\documentclass{elsarticle}

\usepackage[linesnumbered,ruled,vlined]{algorithm2e}
\DontPrintSemicolon

\usepackage{tabularx}
\usepackage{booktabs}
\usepackage{siunitx}

\usepackage{amssymb}
\usepackage{amsmath}
\usepackage{lineno,hyperref}
\usepackage[nameinlink, capitalise]{cleveref}
\usepackage{fullpage}
\usepackage{color}
\usepackage{enumitem}
\usepackage{subcaption}
\usepackage{todonotes}
\usepackage{wrapfig}
\makeatletter
\renewcommand{\todo}[2][]{\@todo[inline,#1]{#2}}
\makeatother

\journal{Journal of Logical and Algebraic Methods in Programming}
\newcommand{\I}{\mathbb{I}}
\newcommand{\Q}{\mathbb{Q}}
\newcommand{\R}{\mathbb{R}}

\newtheorem{definition}{Definition}

\newtheorem{example}{Example}
\def\Def{\mathop{\rm Def}}

\def\Sample{\mathop{\rm Sample}}

\newcommand{\smtrat}{{\small SMT-RAT}\xspace}

\newcommand{\nlsat}{{\small NLSAT}\xspace}
\newcommand{\mcsat}{{\small mcSAT}\xspace}
\newcommand{\nucad}{{\small NuCAD}\xspace}

\newcommand{\fals}{\texttt{False}\xspace}
\newcommand{\tru}{\texttt{True}\xspace}

\definecolor{myPetrol}{RGB}{27,158,119}
\definecolor{myOrange}{RGB}{217,95,2}
\definecolor{myViolet}{RGB}{152,78,163}
\definecolor{myPink}{RGB}{251,154,153}
\definecolor{myGreen}{RGB}{102,166,30}

\begin{document}

\begin{frontmatter}
	\title{Deciding the Consistency of Non-Linear Real Arithmetic Constraints \\ with a Conflict Driven Search Using Cylindrical Algebraic Coverings}
	\author{Erika {\'A}brah{\'a}m\fnref{a}}
        \author{James H. Davenport\fnref{b}}
        \author{Matthew England\fnref{c}}
        \author{Gereon Kremer\fnref{a}}
        \address[a]{RWTH Aachen University, Germany}
        \address[b]{University of Bath, UK}
        \address[c]{Coventry University, UK}
	
\begin{abstract}
We present a new algorithm for determining the satisfiability of conjunctions of non-linear polynomial constraints over the reals, which can be used as a theory solver for satisfiability modulo theory (SMT) solving for non-linear real arithmetic.  
The algorithm is a variant of Cylindrical Algebraic Decomposition (CAD) adapted for satisfiability, where solution candidates (sample points) are constructed incrementally, either until a satisfying sample is found or sufficient samples have been sampled to conclude unsatisfiability.  The choice of samples is guided by the input constraints and  previous conflicts.  

The key idea behind our new approach is to start with a partial sample; demonstrate that it cannot be extended to a full sample; and from the reasons for that rule out a larger space around the partial sample, which build up incrementally into a cylindrical algebraic covering of the space.  There are similarities with the incremental variant of CAD, the NLSAT method of Jovanovi\'{c} and de~Moura, and the NuCAD algorithm of Brown; but we present worked examples and experimental results on a preliminary implementation to demonstrate the differences to these, and the benefits of the new approach.
\end{abstract}
	

\end{frontmatter}

\section{Introduction}
\label{SEC:Intro}

\subsection{Real Algebra}

Formulae in \emph{Real Algebra} are Boolean combinations of polynomial constraints with rational coefficients over real-valued variables, possibly quantified.  Real algebra is a powerful logic suitable to express a wide variety of problems throughout science and engineering.  The 2006 survey \cite{Sturm2006} gives an overview of the scope here.  Recent new applications include bio-chemical network analysis \cite{BDEEGGHKRSW17}, economic reasoning \cite{MDE18}, and artificial intelligence \cite{AMIA14}.  Having methods to analyse real algebraic formulae allows us to better understand those problems, for example, to find one solution, or symbolically describe all possible solutions for them.  

In this paper we restrict ourselves to formulae in which every variable is existentially quantified.  This falls into the field of \emph{Satisfiability Modulo Theories} (\emph{SMT}) solving, which grew from the study of the Boolean SAT problem to encompass other domains for logical formulae.  
In traditional SMT solving, the search for a solution follows two parallel threads: a \emph{Boolean} search tries to satisfy the Boolean structure of $\varphi$, accompanied by a \emph{theory} search that tries to satisfy the polynomial constraints that are assumed to be \tru in the current Boolean search.  To implement such a technique, we need a decision procedure for the theory search that is able to check the \emph{satisfiability of conjunctions of polynomial constraints}, in other words, the \emph{consistency of polynomial constraint sets}.
The development of such methods has been highly challenging. 

Tarski showed that the Quantifier Elimination (QE) problem is decidable for real algebra \cite{Tarski1951}. That means, for each real-algebraic formula $\forall x.\varphi$ or $\exists x.\varphi$ it is possible to construct another, semantically equivalent formula using the same variables as used to express $\varphi$ but without referring to $x$. For conjunctions of polynomial constraints it means that it is possible to decide their satisfiability, and for any satisfiable instance provide satisfying variable values.  Tarski's results were ground breaking, but his constructive solution was non-elementary (with a time complexity greater than all finite towers of powers of $2$) and thus not applicable.

\subsection{Cylindrical Algebraic Decomposition}

An alternative solution was proposed by Collins in 1975 \cite{Collins1975}. Since its invention, Collins' \emph{Cylindrical Algebraic Decomposition} (\emph{CAD}) method was the target of numerous improvements and has been implemented in many Computer Algebra Systems.  Its theoretical complexity is doubly exponential\footnote{Doubly exponential in the number of variables (quantified or free).  The double exponent does reduce by the number of equational constraints in the input \cite{EBD15}, \cite{ED16a}, \cite{EBD19}.  However, the doubly-exponential behaviour is intrinsic: in the sense that classes of examples have been found where the solution requires a doubly exponential number of symbols to write down \cite{BrownDavenport2007}. }.
The fragment of our interest, which excludes quantifier alternation, has lower theoretical complexity of singly exponential time (see for example \cite{BPR06}), however, currently no algorithms are implemented that realise this bound in general (see \cite{Hong1991} for an analysis as to why).   
Current alternatives to the CAD method are efficient but incomplete methods using, e.g., linearisation \cite{cimatti*:linearization,monniaux*:linearization}, interval constraint propagation \cite{isat,Tung2017}, virtual substitution \cite{Article_Weispfenning_Quadratic, CA_FCT11}, subtropical satisfiability \cite{fontaine2017subtropical} and Gr\"obner bases \cite{CoCoALib}. 

Given a formula in real algebra, a CAD may be produced which decomposes real space into a finite number of disjoint cells so that the truth of the formula is invariant within each cell.  Collin's CAD achieved this via a decomposition on which each polynomial in the formula has constant sign. Querying one sample point from each cell can then allow us to determine satisfiability, or perform quantifier elimination.

A full decomposition is often not required and thus savings can be made by adapting the algorithm to terminate early.  For example, once a single satisfying sample point is found we can conclude satisfiability of the formulae\footnote{There are other approaches to avoiding a full decomposition, for example, \cite{WBDE14} suggests how segments of the decomposition of interest can be computed alone based on dimension or presence of a variety. }.  This was first proposed as part of the \emph{partial CAD} method for QE \cite{collins_1991}.   The natural implementation of CAD for SMT performs the decomposition incrementally by polynomial, refining CAD cells by subdivision and querying a sample from each new unsampled cell before performing the next subdivision.

Even if we want to prove unsatisfiability, the decomposition of the state space into sign-invariant cells is not necessary.  There has been a long development in how simplifications can be made to achieve truth-invariance without going as far as sign-invariance such as  \cite{Collins1998a}, \cite{McCallum1999}, \cite{Bradfordetal2016a}.  

\subsection{New Real Algebra Methods Inspired by CAD}

The present paper takes this idea of reducing the work performed by CAD further, proposing that the decomposition need not even be \emph{disjoint}, and neither sign- nor truth-invariant as a whole for the set of constraints.  We will produce cells incrementally, each time starting with a sample point, which if unsatisfying we generalise to a larger cylindrical cell using CAD technology.  We continue, selecting new samples from outside the existing cells until we find either a satisfying sample, or the entire space is covered by our collection of overlapping cells which we call a \emph{Cylindrical Algebraic Covering}.

Our method shares and combines ideas from two other recent CAD inspired methods: 
(1) the \nlsat approach by Jovanovi\'c and de Moura, an instance of the \emph{model constructing satisfiability calculus} (\mcsat) framework \cite{jovanovic2017integer}; and 
(2) the \emph{Non-uniform} CAD (\nucad) approach of Brown \cite{Brown2015}, which is a decomposition of the state space into disjoint cells that are \emph{truth-}invariant for the \emph{conjunction} of a set of polynomial constraints, but with a weaker relationship between cells (the decomposition is not cylindrical).  

We demonstrate later with worked examples how our new approach outperforms a traditional CAD while still differing from the two methods above:  with more effective learning from conflicts than \nucad and, unlike \nlsat, an SMT-compliant approach which keeps theory reasoning separate from the SAT solver. 

\subsection{Paper Structure}

We continue in Section \ref{SEC:Preliminaries} with preliminary definitions and descriptions of the alternative approaches.  We then present our new algorithm, first the intuition in Section \ref{SEC:Intuition} and then formally in Section \ref{SEC:Algo}. Section \ref{SEC:Example} contains illustrative worked examples while Section \ref{SEC:Experiments} describes how our implementation performed on a large dataset from the SMT-LIB~\cite{Barrett2016}.  
We conclude and discuss further research directions in Section \ref{SEC:Conc}.

\section{Preliminaries}
\label{SEC:Preliminaries}

\subsection{Formulae in Real Algebra}

The general problem of solving, in the sense of eliminating quantifiers from, a quantified logical expression which involves polynomial equalities and inequalities with real-valued variables is an old one. 

\begin{definition}%[Quantifier Elimination]
\label{def:QE}
Consider a quantified proposition in \emph{prenex normal form}\index{Prenex normal form}\footnote{Any proposition with quantified variables can be converted into this form --- see any standard logic text.}:
\begin{equation}\label{eq:QE}
Q_1 x_1\ldots Q_{m} x_{m} F(x_1,\ldots,x_{m},x_{m+1},\ldots,x_{n}),
\end{equation}
where each $Q_i$ is either $\exists$ or $\forall$ and $F$ is a semi-algebraic proposition, i.e. a Boolean combination of constraints 
\[
p_j(x_1,\ldots,x_{m},\allowbreak x_{m+1},\ldots,x_{n}) \sigma_j 0,
\]
where $p_j$ are polynomials with rational coefficients and each $\sigma_j$ is an element of $\{<,\le,>,\ge,\ne,=\}$. 

The \emph{Quantifier Elimination} \emph{(QE)} \emph{Problem} is to determine an equivalent quantifier-free semi-algebraic proposition $G(x_{m+1},\ldots,x_{n})$.
\end{definition}

\begin{example}
The formula $\exists y.\, x\cdot y>0$ is equivalent to $x\not= 0$; the formula $\exists y.\, x\cdot y^2>0$ is equivalent to $x>0$; whereas the formula $\forall y.\, x\cdot y^2>0$ is equivalent to \fals.
\end{example}

The existence of a quantifier-free equivalent is known as the \emph{Tarski--Seidenberg Principle}
\cite{Seidenberg1954,Tarski1951}.  The first constructive solution was given by Tarski \cite{Tarski1951}, but the complexity of his solution was indescribable (in the sense that no elementary function could describe it).

\subsection{Cylindrical Algebraic Decomposition}

A better (although doubly exponential) solution had to await the concept of Cylindrical Algebraic Decomposition (CAD) in \cite{Collins1975}.  We start by defining what is meant by an algebraic decomposition here.

\begin{definition}\label{def:AD}~
  \begin{enumerate}

  \item A \emph{cell} from $\R^n$ is a non-empty connected subset of $\R^n$.
    
  \item A \emph{decomposition} of $\R^n$ is a collection $D=\{C_1,\ldots,C_{h}\}$ of finitely many pairwise-disjoint cells from $\R^n$ with $\R^n=\cup_{i=1}^{h} C_i$ .

  \item A cell $C$ from $\R^n$ is \emph{algebraic} if it can be described as the solution set of a formula of the form
\begin{equation}\label{eq:defform}
p_1(x_1,\ldots,x_n)\sigma_10\ \land\ \cdots\ \land\ p_k(x_1,\ldots,x_n) \sigma_k 0,
\end{equation} 
where the $p_i$ are polynomials with rational coefficients and variables from $x_1,\ldots,x_{n}$, and where the $\sigma_i$ are taken from $\{=,>,<\}$\footnote{Since the constraints in (\ref{eq:defform}) need not be equations a more accurate name would be semi-algebraic.  We can avoid $\le$ and $\ge$, since e.g. $p\le0$ is equivalent to $-p>0$.  Avoiding $\ne$ is a more fundamental requirement.}. Equation (\ref{eq:defform}) is a
\emph{defining formula} of $C$, denoted as $\Def(C)$.

  \item A decomposition of $\R^n$ is algebraic if each of its cells is algebraic. 

  \item \label{def:SAD}\index{Algebraic!decomposition!sampled}
A decomposition $D$ of $\R^n$ is \emph{sampled} if it is equipped with a function assigning an explicit point
$\Sample(C)\in C$ to each cell $C\in D$.

  \end{enumerate}
\end{definition}

\begin{example}
\label{exDecomp}
  \begin{itemize}
  \item $D=\{(-\infty,-1),[-1,-1],(-1,1),[1,1],(1,\infty)\}$ is a decomposition of $\R$.
  \item This $D$ is also algebraic because the cells can be described by $x_1<-1$, $x_1=-1$, $x_1>-1\wedge x_1<1$, $x_1=1$ and $x_1>1$, respectively, in their order of listing above.
  \item To get a sampled algebraic decomposition, we additionally provide $-2$, $-1$, $\tfrac{1}{2}$, $1$ and $3$, respectively.%, again for the cell order above.
%    \item This decomposition $D$ is sign-invariant for $x_1^2-1$.
  \end{itemize}
\end{example}

We are interested in decompositions with certain important properties relative to a set of polynomials as formalised in the next definition.

\begin{definition}\label{def:SID}
A cell $C$ from $\R^n$ is said to be \emph{sign-invariant} for a
polynomial $p(x_1,\ldots,x_n)$ if and only if
precisely one of the following is \texttt{True}:
\[
(1) \quad \forall {\bf x}\in C.\, p({\bf x})>0; \mbox{ or } \qquad
(2) \quad \forall {\bf x}\in C.\, p({\bf x})<0; \mbox{ or } \qquad
(3) \quad \forall {\bf x}\in C.\, p({\bf x})=0.
\]
%\begin{enumerate}
%\item $\forall {\bf x}\in C\quad p({\bf x})>0$; or 
%\item $\forall {\bf x}\in C\quad p({\bf x})<0$; or
%\item $\forall {\bf x}\in C\quad p({\bf x})=0$.
%\end{enumerate}

A cell $C$ is sign-invariant for a set of polynomials if and only if it is sign-invariant for each polynomial in the set individually.
A decomposition of $\R^n$ is sign-invariant for a polynomial (a set of polynomials) if each of its cells is sign-invariant for the polynomial (the set of polynomials).
\end{definition}
For example, the decomposition in Example \ref{exDecomp} is sign-invariant for $x_1^2-1$.

All of the techniques discussed in this paper to produce decompositions are defined with respect to an ordering on the variables in the formulae.

\begin{definition}
\label{def:ordering}
For positive natural numbers $m<n$, we see $\R^n$ as an extension of $\R^m$ by further dimensions, and denote the coordinates of $\R^m$ as $x_1,\ldots,x_m$ and the coordinates of $\R^n$ as $x_1,\ldots,x_n$. 
Unless specified otherwise we assume polynomials in this paper are defined with these variables under the ordering corresponding to their labels, i.e., $x_1 \prec x_2 \prec \dots \prec x_n$.  

We define the \emph{main variable} of a polynomial as the highest variable in the ordering that is present in the polynomial.  If we say that a set of polynomials are in $\R^{i}$ then we mean that they are defined with variables $x_1, \dots, x_i$ and at least one such polynomial has main variable $x_i$.
\end{definition}

We can now define the structure of the cells in our decomposition.  

\begin{definition}\label{def:globalCyl}
Let $m<n$ be positive natural numbers. A decomposition $D$ of $\R^n$ is said to be \emph{cylindrical} 
over a decomposition $D'$ of $\R^m$
if the projection onto $\R^m$ of every cell of $D$ is a cell of $D'$. I.e. the projections of any pair of cells from $D$ are either identical (if the same cell in $D'$) or disjoint (if different ones).

The cells of $D$ which project to $C\in D'$ are said to form the \emph{cylinder}\index{Cylinder}
over $C$.  %, denoted $\Cyl(C)$. 
For a sampled decomposition, we also
insist that the sample point in $C$ be the projection of the sample points of
all the cells in the cylinder over $C$. %\rm This definition is usually stated when $m=n-1$, but the greater generality is theoretically worth having, even though we only currently know  how to compute these when $m=n-1$.
A (sampled) decomposition $D$ of $\R^n$ is cylindrical if and only if for each positive natural number $m<n$ there exists a (sampled) decomposition $D'$ of $\R^m$ over which $D$ is cylindrical.
\end{definition}

\begin{wrapfigure}[9]{r}{6cm}
\includegraphics{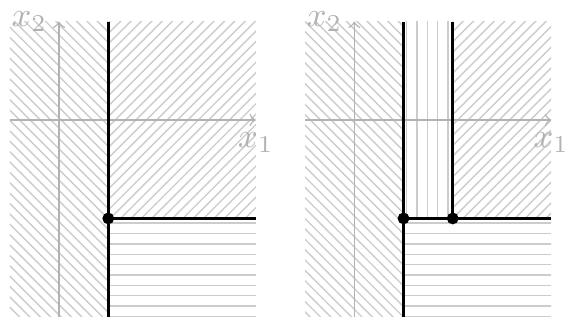}
  \caption{Illustrations of cylindricity.}
    \label{fig:cylindrical}
  \end{wrapfigure}

So combining the above definitions we have a (Sampled) Cylindrical Algebraic Decomposition (CAD).  Note that cylindricity implies the following structure on a decomposition cell.

\begin{definition}\label{def:localCyl}
A cell from $\R^n$ is \emph{locally cylindrical} if it can be described by conditions $c_1(x_1)$, $c_2(x_1, x_2)$, $\dots$, $c_n(x_1, \dots, x_n)$ where each $c_i$ is one of:
\begin{align*}
\ell_i(x_1, \dots, x_{i-1}) < &x_i, \\
\ell_i(x_1, \dots, x_{i-1}) < &x_i < u_i(x_1, \dots, x_{i-1}),\\
&x_i < u_i(x_1, \dots, x_{i-1}),  \\
&x_i = s_i(x_1, \dots, x_{i-1}). 
\end{align*} 
Here $\ell_i, u_i, s_i$ are functions in $i-1$ variables (constants when $i=1$). 
These functions can be rational polynomials or indexed root expressions (whose values for given $x_i,\dots,x_{n}$ might be algebraic numbers).

\end{definition}

\begin{example}
\cref{fig:cylindrical} shows two decompositions of $\R^2$ in which, each dot, each line segment, and each hatched region are cells. The decomposition of $\R^2$ on the left of the figure is cylindrical over $\R^1$ (horizontal axis), but the decomposition on the right is not (two cells have overlapping non-identical projections onto $x_1$). All cells are locally cylindrical. Note that we are assuming $x_1 \prec x_2$; for $x_2 \prec x_1$ neither of the decompositions would be cylindrical, but still each cell would be locally cylindrical.
\end{example}

Collins' solution to the QE problem in \cref{def:QE} was an algorithm to produce a CAD \cite{Collins1975} of $\R^{n}$, sign-invariant for all the $p_j$, and thus truth-invariant for $F$.  Hence it is then only necessary to consider the truth or falsity of $F$ at the finite number of sample points and query their algebraic definitions to form $G$.  Furthermore, if $Q_i$ is $\forall$, we require $F$ to be \tru at all sample points, whereas $\exists$ requires the truth of $F$ for at least one sample point.  
As has been pointed out \cite{Collins1998a,Bradfordetal2016a}, such a CAD is finer than required for QE since as well as answering the QE question asked, it could answer another that involved the same $p_j$; but possibly different $\sigma_j$, and even different $Q_i$ as long as the variables are quantified in the same order.

\subsection{Our Setting: SMT for Real Algebra}
\label{SSEC:Setting}

In this paper we are interested in the subset of QE problems coming from \emph{Satisfiability Modulo Theories} (SMT) solving \cite{BSST09}, namely sentences (formulae whose variables are all bound by quantifiers) which use only the existential quantifier.  I.e. (\ref{eq:QE}) with $m=n$ and all $Q_i = \exists$.  

Traditional SMT-solving aims to decide the truth of such formulae by searching for solutions satisfying the Boolean structure of the problem using a SAT-solver, and concurrently checking the consistency of the corresponding theory constraints.  It additionally restricts to the case that $F$ is a pure conjunction of the $p_j\sigma_j0$; and makes the following additional requirements on any solution:
\begin{enumerate}
\item If the answer is \tru (generally referred to as SATisfiable), we need only an explicit point at which it is \tru (rather than a description of all such points);
\item If the answer is \fals (generally referred to as UNSATisfiable), we want some kind of justification that can be used to guide the SAT-solver's search.
\end{enumerate}
So, in SMT solving we are less interested in complete algebraic structures but rather in deciding satisfiability and computing solutions if they exist. Reflecting this interest, Real Algebra is often called Non-Linear Real Arithmetic (NRA) in the SMT community. The use of CAD (and computer algebra tools more generally) in SMT has seen increased interest lately \cite{Abraham2015b, AAB+16a} and several SMT-solvers make use of these \cite{smtrat, FOSKT18}.  

\subsection{Relevant Prior Work}
\label{sec:related}

Our contribution is a new approach to adapt CAD %tools: JHD we aren't really using any CAD tools as such
technology and theory for this particular problem class.  There are three previous works that have also sought to adapt CAD ideas to the SMT context:
\begin{itemize}
\item \textbf{Incremental CAD} is an adaptation of traditional CAD so that it works incrementally, checking the consistency of an increasing number of polynomial constraints, as implemented in the \smtrat solver~\cite{smtrat,Kremer2019}.  Traditional CAD as formulated by Collins and implemented in Computer Algebra Systems consists of two sequentially combined phases (projection and lifting): it will first perform all projection computations to produce algebraic data, and then use this to construct all the cells (with sample points).  The incremental adaptation instead processes one projection operation at a time, and then uses that to derive any additional cells (with their additional samples), before making the next projection computation.  For the task of satisfiability checking, this allows for early termination not only for satisfiable problems (if we find a sample that satisfies all sign conditions then we do not need to compute any remaining projections) but also for unsatisfiable ones (if the CAD for a subset of the input constraints is computed and no sample from its cells satisfies all those sign conditions).  Although the implementation is technically involved, the underlying theory is traditional CAD and in the case of UNSAT that is exactly what is produced.  
  
\item \textbf{The \nlsat Algorithm} by Jovanovi\'c and de Moura \cite{JovanovicdeMoura2012a} lifts the theory search to the level of the Boolean search, where the search at the Boolean and theory levels are mutually guided by each other away from unsatisfiable regions when it can be determined by some kind of propagation or lookahead. Partial solution candidates for the Boolean structure and for the corresponding theory constraints are constructed incrementally in parallel until either a full solution is found or a conflict is detected, meaning that the candidates cannot be extended to full solutions. Boolean conflicts are generalised using propositional resolution. For theory conflicts, CAD technology is used to generalise the conflict to an unsatisfying region (a CAD cell). These generalisations are learnt by adding new clauses that exclude similar situations from further search by the above-mentioned propagation mechanisms.  
In the case of a theory conflict, the incremental construction of theory solution candidates allows to identify a minimal set of constraints that are inconsistent under the current partial assignment. This minimal constraint set induces a coarser state space decomposition and thus in general larger cells that can be used to generalise conflicts and exclude from further search by learning. The exclusion of such a cell is learnt by adding a new clause expressing that the constraints are not compatible with the (algebraic description of the) unsatisfying cell. This clause will lead away from the given cell not only locally but for the whole future search when the constraints are all \tru.

\item \textbf{The Non-Uniform CAD (\nucad) Algorithm} by Brown \cite{Brown2015} takes as input a set of polynomial constraints and computes a decomposition whose cells are \emph{truth-invariant} for the \emph{conjunction} of all input constraints. It starts with the coarsest decomposition, having the whole space as one cell, which does not guarantee any truth invariance yet.  We split it to smaller cells that are sign-invariant for the polynomial of one of the input constraints, and mark all refined cells that violate the chosen constraint as final: they violate the conjunction and are thus truth-invariant for it. Each refinement is made by choosing a sample in a non-final cell and generalising to a locally cylindrical cell.  At any time all cells are locally cylindrical, but there is no global cylindricity condition.  The algorithm proceeds with iterative refinement, until all cells are marked as final.  Each refinement will be made with respect to one constraint for which the given cell is not yet sign-invariant.  
There are two kinds of cells in a final \nucad: cells that violate one of the input constraints (but these cells are not necessarily sign- nor truth-invariant for all other constraints), and cells that satisfy all constraints (and are sign-invariant for all of them).  The resulting decomposition is in general coarser (i.e. it has less cells) than a regular CAD. Its cells are neither sign- nor truth-invariant for individual constraints, but they are truth-invariant for their conjunction and that is sufficient for consistency checking.

\end{itemize}

\section{Intuitive Idea Behind Our New Algorithm}
\label{SEC:Intuition}

\subsection{From Sample to Cell in Increasing Dimensions}

We want to check the consistency of a set of input polynomial sign constraints, i.e., the satisfiability of their conjunction. Traditional CAD first generates  algebraic information on the formula we study (the projection polynomials) and then uses these to construct a set of sample points, each of which represents a cell in the decomposition\footnote{An incremental CAD approach would incrementally create new projection polynomials and new sample points.}.  Our new approach works the other way around: we will select (or guess) a sample point by fixing it dimension-wise, starting with the lowest dimension and iteratively moving up in order to extend lower-dimensional samples to higher-dimensional ones. Thus we start with a zero-dimensional sample and iteratively explore new dimensions by recursively executing the following:
\begin{itemize}

\item Given an $(i{-}1)$-dimensional sample $s=(s_1,\ldots,s_{i{-}1})$ that does not evaluate any input constraint to \fals, we extend it to a sample $(s_1,\ldots,s_{i{-}1},s_i)$, which we denote as $s\times s_i$.

\item If this $i$-dimensional sample can be extended to a satisfying solution, either by recursing or because it already has full dimension, then we terminate, reporting consistency (and this witness point).

\item Otherwise we take note of the reason (data on the conflicting requirements that explains why the sample can not be extended) and exclude from further search not just this particular sample $s\times s_i$, but all extensions of $s$ into the $i$th dimension with any value from a (hopefully large) interval $I$ around $s_i$ which is unsatisfiable for the same reason.

\end{itemize}
This means that for future exploration, the algorithm is guided to look somewhere away from the reasons of previous conflicts.  This should allow us to find a satisfying sample quicker, or in the case of UNSAT build a covering of all $\R^n$ such that we can conclude UNSAT everywhere with fewer cells than a traditional CAD.  
%The important part is to make these characterization composable: once we excluded certain areas of the solution space because they violated some input constraints, the next exclusions may be based on these existing ones (as well as the input constraints), as detailed in the following.
In the last item above, the generalisation of an unsatisfying sample $s\times s_i$  to an unsatisfying interval $s\times I$ will use a constraint $p(x_1,\ldots,x_i)\sigma 0$ that is violated by the sample, i.e., $p(s\times s_i)\sigma 0$ does not hold. 
It might be the case that $s_i$ is a real zero of $p(s)$ and we then have $I=[s_i,s_i]$.  Otherwise we get an interval $I=(\ell,u)$ with $\ell<s_i<u$ where $\ell$ is either the largest real root of $p$ below $s_i$ or $-\infty$ if no such root exists (and analogously $u$ is either the smallest real root above $s_i$ or $\infty$). Thus we have $p(s\times r)\not=0$ for all $r\in (\ell,u)$. Since $p(s\times s_i)\sigma 0$ is \fals and the sign of a polynomial does not change between neighbouring zeros, we can safely conclude that $p(x_1,\ldots,x_i)\sigma 0$ is violated by all samples $(s\times r)$ with $r\in I$.

We continue and check further extensions of $s=(s_1,\ldots,s_{i{-}1})$, until either we find a solution or the $i$th dimension is fully covered by excluding intervals. In the latter case, we take the collection of intervals produced and use CAD projection theory to rule out not just the original sample in $\mathbb{R}^{i{-}1}$ but an interval around it within dimension $(i{-}1)$, i.e. the same procedure in the lower dimension. Intuitively, each sample $s\times s_i$ violating a constraint with polynomial $p$ can be generalised to a cell in a $p$-sign-invariant CAD. So when all extensions of  $s$ have been excluded (the $i$th dimension is fully covered by excluding intervals) then we project all the covering cells to dimension $i{-}1$ and exclude their intersection from further search.

\subsection{Restoring Cylindricity}

If we were to generalise violating samples to cells in a CAD that is sign-invariant for \emph{all} input constraints then in the case of unsatisfiability we would explore a CAD structure. But instead, by \emph{guessing} samples in not yet explored areas and identifying violated constraints individually per sample, we are able to generalise samples to larger cells that can be excluded from further search: like in the \nlsat approach.  

Unlike \nlsat, we then build the \emph{intersection} creating a cylindrical arrangement at the cost of making the generalisations smaller (but as  large as possible while still ordered cylindrically).
What is the advantage gained from a cylindrical ordering?  In \nlsat the excluded cells are not cylindrically ordered; and so SMT-mechanisms are used in the Boolean solver (like clause learning and propagation) to lead the search away from previously excluded cells. In contrast, our aim is to make this book-keeping remain inside the algebraic  procedure, which can be done in a depth-first-search approach when the cells are cylindrically ordered. 

\subsection{Cylindrical Algebraic Coverings}

So, we maintain cylindricity from CAD, but we relax the disjointness condition on cells in a decomposition, allowing our cells to \emph{overlap} as long as their cylindrical ordering is still maintained. Instead of decomposition we will use the name \emph{covering} for these structures.

%\begin{definition}
%\label{def:cover}
%Given a set $P$ of polynomials in $n$ variables and a sample $s\in\R^i$, $0\leq d<n$, a \emph{algebraic covering for $P$ and $s$} is a collection of intervals whose bounds are defined as indexed root expressions\footnote{i.e. unique roots of multivariate polynomials over the sample point $s$ (possibly involving algebraic numbers)} that together cover $\R$.
%where the real line concerned is considered as the dimension above a fixed sample point.  
%
%An \emph{UNSAT covering for $P$ and $s$} is an algebraic covering for $P$ and $s$ in which every interval has been fully explored and no extensions for $s$ were found that satisfy $P$.
%\end{definition}

\begin{definition}\label{def:cover}~
  \begin{enumerate}

  \item A \emph{covering} of $\R^n$ is a collection $D=\{C_1,\ldots,C_h\}$ of finitely many (not necessarily pairwise-disjoint) cells from $\R^n$ with $\R^n=\cup_{i=1}^h C_i$ .

  \item A covering of $\R^n$ is \emph{algebraic} if each of its cells is algebraic. 

\item
A covering $D$ of $\R^n$ is \emph{sampled} if it is equipped with a function assigning an explicit point
$\Sample(C)\in C$ to each cell $C\in D$.

\item A cell $C$ from $\R^n$ is \emph{UNSAT} for a
polynomial constraint (set of constraints) if and only if all points in $C$ evaluate the constraint (at least one of the constraints) to \fals.
A covering of $\R^n$ is UNSAT for a constraint (set) if each of its cells is UNSAT for the constraint (at least one from the set).

\item
  A covering $D$ of $\R^n$ is said to be \emph{cylindrical} over a
  covering $D'$ of $\R^m$ if the projection onto $\R^m$ of every cell
  of $D$ is a cell of $D'$.  The cells of $D$ which project to $C\in
  D'$ form the \emph{cylinder} % $\Cyl(C)$ 
  over $C$. For a sampled
  covering, the sample point in $C$ needs to be the
  projection of the sample points of all the cells in the cylinder
  over $C$.
A (sampled) covering $D$ of $\R^n$ is cylindrical if and only if for each $0<m<n$ there exists a (sampled) covering $D'$ of $\R^m$ over which $D$ is cylindrical.

  \end{enumerate}
  
\end{definition}

The coverings produced in our algorithm are all UNSAT coverings for constraint sets (i.e. at least one constraint is unsatisfied on every cell).  %In future work we will consider how this may be generalised to formulae and to mixed SAT/UNSAT coverings.

%From such a covering we produce a \emph{justification}: a set of polynomials that characterizes the boundaries of this UNSAT covering in the lower dimensions.  This allows for the cover to be expanded beyond the sample point.  Specifically, it is \emph{valid} on the region around the sample point for which the justification is sign-invariant. The components of the justification essentially encode different possibilities for when the UNSAT covering \emph{ceases to be an UNSAT covering}.

%The full algorithm roughly works as follows: given a partial sample point (that does not directly conflict with any input constraints) we try to extend it in two steps: firstly we identify all areas in the next dimension that immediately conflict with an input constraint and secondly recurse with guessed extensions  until no further extension is possible.

\subsection{Differences to Existing Methods}

Our new method shares and combines ideas from the related work in Section \ref{sec:related} but differs from each.
\begin{itemize}

\item  A version of the CAD method which proceeds incrementally by constraint or projection factor is implemented in the \smtrat solver.  Our new method differs as even in the case of unsatisfiability it will not need to compute a full CAD but rather a smaller number of potentially overlapping cells.
  
\item While its learning mechanism made \nlsat the currently most successful SMT solution for Real Algebra, it brings new scalability problems by the large number of learnt clauses. Our method is conflict-driven like \nlsat, but instead of learning clauses, we embed the learning in the algebraic procedure. Learning clauses allows to exclude unsatisfiable CAD cells for certain combinations of polynomial constraints for the whole future search, but it also brings additional costs for maintaining the truth of these clauses. Our approach remembers the unsatisfying nature of cells only for the current search path, and learns at the Boolean level only unsatisfiable combinations of constraints. To unite the advantages from both worlds, our approach could be extended to a hybrid setting where we learn at both levels (by returning information on selected cells to be learned as clauses).
  
\item Like \nucad, our algorithm can compute refinements according to different polynomials in different areas of the state space and to stop the refinement if any constraint is violated, however we retain the global cylindricity of the decomposition. Furthermore, driven by model construction, we can identify minimal sets of relevant constraints that we use for cell refinement, instead of the arbitrary choice of these polynomials used by \nucad. Our expectation is thus that on average this will lead to coarser decompositions, and certainly rule out some unnecessary worst case decompositions.  

\end{itemize}

\section{The New Algorithm}
\label{SEC:Algo}

\subsection{Interface, I/O, and Data Structure}
\label{SSEC:IO}

\begin{algorithm}[b]
	\SetKwInOut{Input}{Input}\SetKwInOut{Output}{Output}
	\KwData{Global set $C$ of polynomial constraints defined over $\R^n$.}
	\Input{None}
	\Output{Either $(\mathrm{SAT}, S)$ where $S \in \R^n$ is a full-dimensional satisfying witness; \\
		or $(\mathrm{UNSAT}, \overline{C})$ when no such $S$ exists, where $\overline{C} \subset C$ is also unsatisfiable. 
	}
	($\textit{flag}$, $\textit{data}$) := {\texttt{get$\_$unsat$\_$cover}( () )}\tcp*{Algorithm \ref{alg:main}} 
	\eIf{$\textit{flag} = \mathrm{SAT}$}{ \Return{ (flag, data) }}{ \Return{ (flag, \emph{\texttt{infeasible$\_$subset}}(data)) \nllabel{line:unsatcore}} }
	\caption{\texttt{user$\_$call()} \label{alg:user}}
\end{algorithm}

\begin{algorithm}[t]
	\SetKwInOut{Input}{Input}\SetKwInOut{Output}{Output}
	\KwData{Global set $C$ of polynomial constraints defined over $\R^n$.}
	\Input{Sample point $s=(s_1,\ldots,s_{i-1}) \in \R^{i-1}$ such that no global constraint evaluated at $s$ is \fals.  Note that if $s =()$ then $i=1$ (i.e. we start from the first dimension).}
	\Output{Either $(\mathrm{SAT}, S)$ where $S \in \R^n$ is a full-dimensional satisfying witness; \\
		or $(\mathrm{UNSAT}, \I)$ when $s$ cannot be extended to a satisfying sample in $\R^n$.  $\I$ represents a set \\ of intervals which cover $s \times \R$ and come with algebraic information (see Section \ref{SSEC:IO}). 
	}

	$\I :=$ \texttt{get$\_$unsat$\_$intervals(s)}\nllabel{unsatintervals}\tcp*{Algorithm \ref{alg:queryC}}
	\While{$\bigcup_{I \in \I} I \neq \R$}{
		$s_i :=$ \texttt{sample$\_$outside}$(\I)$ \nllabel{s2} \;
		\If{$i=n$}{
			\Return{\nllabel{retSAT1}$(\mathrm{SAT}, (s_1,\ldots,s_{i-1},s_i))$}
		}
		$(f,O) :=$ \texttt{get$\_$unsat$\_$cover($(s_1,\ldots,s_{i-1},s_i)$)} \nllabel{recurse} \tcp*{recursive call}
		\uIf(\tcp*[f]{then $O$ is a satisfying sample}){$f = \mathrm{SAT}$ }{
			\Return{\nllabel{retSAT2}$(\mathrm{SAT},O)$ \tcp*{pass answer up to main call}}
		} 
		\uElseIf(\tcp*[f]{then $O$ is an unsat covering}){$f = \mathrm{UNSAT}$ }
		{
			$R :=$ \texttt{construct$\_$characterization($(s_1,\ldots,s_{i-1},s_i), O$)} \nllabel{algCJ} \tcp*{Algorithm \ref{alg:construct-characterization}}
			$I :=$ \texttt{interval$\_$from$\_$characterization($(s_1,\ldots,s_{i-1}), s_i, R$)} \nllabel{algIFJ} \tcp*{Algorithm \ref{alg:interval-from-characterization}}
			$\I := \I \cup \{I\}$ \nllabel{e2}\;
		}
	}
	\Return{$(\mathrm{UNSAT}, \I)$ \label{line:finalRet}}
	
	\caption{\texttt{get$\_$unsat$\_$cover($s$)} \label{alg:main}}
\end{algorithm}

Our main algorithm, \texttt{get\_unsat\_cover}, is presented as Algorithm \ref{alg:main}, while Algorithm \ref{alg:user} provides the user interface to it.  A user is expected to define the set of constraints whose satisfiability we want to study globally, and then make a call to Algorithm \ref{alg:user} with no input.  This will then call the main algorithm with an empty tuple as the initial sample $s$; the main algorithm is recursive and will later call itself with non-empty input. In these recursive calls the input is a partial sample point $s \in \R^{i{-}1}$ which does not evaluate any global constraint defined over $\R^{i{-}1}$ to \fals, and for which we want to explore dimension $i$ and above.  

The main algorithm provides two outputs, starting with a flag.  When the flag is SAT then the partial sample $s$ was extended to a full sample from $\R^n$ (the second output) which satisfies the global constraints.  
When the flag is UNSAT then the method has explored the higher dimensions and determined that the sample cannot be extended to a satisfying solution.  It does this by computing an \emph{UNSAT cylindrical algebraic covering} for the constraints with the partial sample $s$ substituted for the first $i{-}1$ variables.  Information on the covering, and the projections of these cells, are all stored in the second output in this case.  

More formally, the output $\I$ is a set of objects $I$ each of which represent an interval of the real line (specifically $s \times \R$). We use $I$ later to mean both the interval, and our data structure encoding it which carries additional algebraic information.  In total such a data structure $I$ has six attributes, starting with:
\begin{itemize}[noitemsep,topsep=0pt,parsep=0pt,partopsep=0pt]
\item the lower bound $\ell$;
\item the upper bound $u$;
\item a set of polynomials $L$ that defined $\ell$;
\item a set of polynomials $U$ that defined $u$.
\end{itemize}
The bounds are constant, but potentially algebraic, numbers.  The polynomials define them in that they are multivariate polynomials which when evaluated at $s$ became univariate with the bound as a real root.   The final two attributes are also sets of polynomials:
\begin{itemize}[noitemsep,topsep=0pt,parsep=0pt,partopsep=0pt]
\item a set of polynomials $P_i$ (multivariate with main variable $x_i$);
\item a set of polynomials $P_\bot$ (multivariate with main variable smaller than $x_i$).
\end{itemize} 
These polynomials have the property that allows for generalisation of $s$ to an interval: the property is that perturbations of $s$ which do not change the signs of these polynomials will result in the interval $I$ (whose numerical bounds may have also perturbed but will still be defined by the same ordered real roots of the same polynomials) remaining a region of unsatisfiability.  

Within a covering there must also be special intervals which run to $\infty$ and $-\infty$.  For intervals with these bounds we store the polynomials from the constraints which allowed us to conclude the infinite interval.  

In the case of UNSAT, the user algorithm will have to process the data $\I$ into an \emph{infeasible subset} (Algorithm \ref{alg:user} Line \ref{line:unsatcore}), i.e. a subset of the constraints that are still unsatisfiable.  Ideally this would be minimal (a minimal infeasible subset) although any reduction of the full set would carry benefits.  We discuss how we implement this later in Section \ref{SUBSEC:TheorySolver}.  We note that the correctness of Algorithm \ref{alg:user} follows directly from the correctness of its sub-algorithms.

\subsection{Initial Constraint Processing}

The first task in Algorithm \ref{alg:main} is to see what effect the partial sample $s$ has on the global constraints.  This is described as Algorithm \ref{alg:queryC}, which will produce those intervals $I \subseteq \R$ such that $s \times I$ is conflicting with some input constraints (a partial covering). This method resembles a CAD lifting phase where we substitute a sample point into a polynomial to render it univariate, compute the real roots, and decompose the real line into sign-invariant regions.  Here we do the same for the truth of our input constraints.

\begin{algorithm}[t]
	\SetKwInOut{Input}{Input}\SetKwInOut{Output}{Output}
	\KwData{Global set $C$ of polynomial constraints defined over $\R^n$.}
	\Input{Sample point $s$ of dimension $i-1$ with $i$ a positive integer ($s$ is empty for $i=1$).}
	\Output{$\I$ which represents a set of unsatisfiable intervals over $s \times I$ and comes with some algebraic information (see Section \ref{SSEC:IO}).}  
	$\I := \emptyset$ \;
	$C_i:=$ set of all constraints from $C$ with main variable $x_i$\;% or higher\;
	\ForEach(\tcp*[f]{I.e. $p$ is the defining polynomial of $c$}){constraint $c = p \sigma 0$ in $C_i$}{
		$c' := c(s)$ \tcp*{Substitute $s$ into $c$ to leave univariate}
		\If{\nllabel{const1}$c' = \fals$ }{
		    Define $I$ with $I_\ell = -\infty, I_u = \infty, I_L=\emptyset, I_U=\emptyset, I_{P_i}=\{p\}, I_{P_\bot}=\emptyset$\;
			\Return{$\{I\}$ \nllabel{GSIret1}
			%$\{(-\infty, \infty, \emptyset, \emptyset, \{p\}, \emptyset)\}$
			}
		}
		\If{$c' = \tru$}{
			\texttt{continue} to next constraint\nllabel{const2}
		}
		\tcp{at this point $c'$ is univariate in the $i$th variable}
			$Z :=$ \texttt{real$\_$roots($p, s$)} \tcp*{$Z = [z_1, \dots, z_k]$}
			$Regions := \{ (-\infty,z_1), [z_1,z_1], (z_1, z_2), ..., (z_k,\infty) \}$ \tcp*{$\{(-\infty, +\infty)\}$ if $Z$ was empty}
			\ForEach{$I \in Regions$}{
			   Let $\ell$ and $u$ be the lower and upper bounds of $I$\;
			   Pick $r\in I$ \tcp*{e.g. $r:=\ell+(u-\ell)/2$}
				\If{$c'(r) = \fals$}{
					Set $L,U$ each to $\emptyset$ \;
					\lIf{$\ell \neq -\infty$}{$L := \{p\}$}
					\lIf{$u \neq \infty$}{$U := \{p\}$}
					Define new interval $I$ with $I_\ell = \ell, I_u = u, I_L=L, I_U=U, I_{P_i}=\{p\}, I_{P_\bot}=\emptyset$ \tcp*{Polynomials stored here undergo simplification $-$ see Section \ref{SSEC:PolynomialBases}}
					Add $I$ to $\I$\;
				}
			}
		}
	\Return{\nllabel{GSIret2}$\I$}
	
	\caption{\texttt{get$\_$unsat$\_$intervals($s$)} \label{alg:queryC}}
\end{algorithm}

Algorithm \ref{alg:queryC} Lines \ref{const1}$-$\ref{const2} deal with the case where after substitution the truth value of the constraint may be immediately determined (e.g. the defining polynomial evaluated to a constant).  The constraint either provides the entire line as an UNSAT interval, or no portion of it.  The rest of the code deals with the case where the substitution rendered a constraint univariate in the $i$th variable.  We use \texttt{real$\_$roots(p, s)} to return all real roots of a polynomial $p$ over a partial sample point $s$ that sets all but one of its variables.  We do not specify the details of the real root isolation algorithm here\footnote{Our implementation in SMT-RAT uses bisection with Descartes' rule of signs.} but note that it will need to handle potentially algebraic coefficients.  We assume roots are returned in ascending numerical order with any multiple roots represented as a single list entry. 

The inner for loop queries a sample point in each region of the corresponding decomposition of the line to determine any infeasible regions for the constraint, storing them in the output data structure $\I$.  $\I$ represents a set of intervals $I$ such that $s \times I$ conflicts with some input constraint.  The intervals from $\I$ may be overlapping, and some may be redundant (i.e. fully contained in others).  We discuss this issue of redundancy further in Section \ref{ssec:redundancy}.

It is clear that Algorithm \ref{alg:queryC} will meet its specification, in that it will define intervals on which constraints are unsatisfiable:   the falsity of a constraint caused the inclusion of an interval in the output while the bounds of the interval were defined to ensure that the polynomial defining the failing constraint would not change sign inside.  The role and property of the stored algebraic information will be discussed in Section \ref{sec:alggen}.

The call to Algorithm \ref{alg:queryC} initialises $\I$ in Algorithm \ref{alg:main} in which we will build our UNSAT covering.  It is unlikely but possible that $\I$ already covers $\R$ after the call to Algorithm \ref{alg:queryC}: it could even contain $(-\infty,\infty)$.  If $\I$ is already a covering then we would skip the main loop of Algorithm \ref{alg:main} and directly return it.  For example, this would be triggered by either of the constraints $y^2x<0$ or $y^2 + x < 0$ at sample $s = (x \mapsto 0)$.  The former would have been returned early by \cref{GSIret1} of Algorithm \ref{alg:queryC} while the latter would have required real root isolation and have been returned in \cref{GSIret2} of Algorithm \ref{alg:queryC}.

\subsection{The Main Loop of Algorithm \ref{alg:main}}

We will iterate through Lines $\ref{s2}-\ref{e2}$ of Algorithm \ref{alg:main} until the set of intervals represented by $\I$ cover all $\R$.  In each iteration we collect additional intervals for our UNSAT covering.  

To do this we first generate a sample point $s_i$ from $\R \setminus (\cup_{I\in\I}I)$ using a subroutine \texttt{sample$\_$outside($\I$)} which is left unspecified.  This could simply pick the mid-point in any current gap, or perhaps something more sophisticated (a common strategy would be to prefer integers, or at least rationals).  

Note that $s \times s_i$ necessarily satisfies all those constraints with main variable $x_i$, otherwise we would have generated an interval excluding $s_i$ at \cref{unsatintervals}, as well as all constraints with smaller main variables (from the input specification on $s$).  This means that: (a) if $s \times s_i$ has full dimension, we have found a satisfying sample point for the whole set of constraints and can return this along with the SAT flag in \cref{retSAT1}; and (b) if not full dimension then we will meet the input specification for the recursive call on \cref{recurse}.  The recursion means we will explore $s \times s_i$ in the next dimension up.  The previous check on dimension acts as the base case and thus the recursion is clearly bounded in depth, ensuring termination if the main loop always terminates.

If the result of the recursive call is SAT we simply pass on the result in \cref{retSAT2}.  Otherwise we have an UNSAT covering for $s \times s_i$ and our next task is to see whether we can generalise this knowledge to rule out not just $s_i$, but an interval around it.  We do this in two steps.  
\begin{itemize}
\item First in \cref{algCJ} we call Algorithm \ref{alg:construct-characterization} to construct a \emph{characterisation} from the UNSAT covering: a set of polynomials which were used to determine unsatisfiability and with the property that on the region around $s \times s_i$ where none of them change sign the reasons for unsatisfiability generalise.  I.e., while the exact bounds of the intervals in the coverings may vary: (a) they are still defined by the same ordered roots of the same polynomials (over the sample); and (b) they do not move to the extent that the line is no longer covered.
\item Second in \cref{algIFJ} we call Algorithm \ref{alg:interval-from-characterization} to find the interval in dimension $i$ over $s$ in which those characterisation polynomials are sign-invariant.
\end{itemize}
We describe these two sub-algorithms in detail in the next subsection.  

The interval produced by Algorithm \ref{alg:interval-from-characterization} may be $(-\infty,\infty)$. In this case all other intervals in $\I$ are now redundant and the main loop of Algorithm \ref{alg:main} stops.  Otherwise we continue to iterate.
 
The loop will terminate because although the search space is infinite the combinations of constraints is not.  Each generalisation rules out a portion of space defined by a set of polynomials all having invariant sign on it.  The number of polynomials computed is finite and their changes in sign are finite.  Thus eventually we must either cover all the line or sample in a satisfying region.  This termination argument is very similar to traditional CAD but here we aim to compute fewer, larger overlapping regions.

The correctness of Algorithm \ref{alg:main} thus depends on the correctness of these two crucial sub-algorithms. 

%Note that we may have missed the loop entirely if the whole line were covered by the output of step \ref{unsatintervals}.  In either case if we reach \cref{line:finalRet} the data structure $\I$ represents an UNSAT covering, which is returned, either to the parent recursive call or to the user.

\subsection{Generalising the UNSAT Covering from the Sample}
\label{sec:alggen}

It remains to examine the details of how the UNSAT covering is generalised from the sample $s$ to an interval around it (the calls to Algorithms \ref{alg:construct-characterization} and \ref{alg:interval-from-characterization} on Lines \ref{algCJ} and \ref{algIFJ} of Algorithm \ref{alg:main}).

\subsubsection{Ordering within a covering}
\label{SSEC:Ordering}

\begin{figure}
	\centering
        \includegraphics{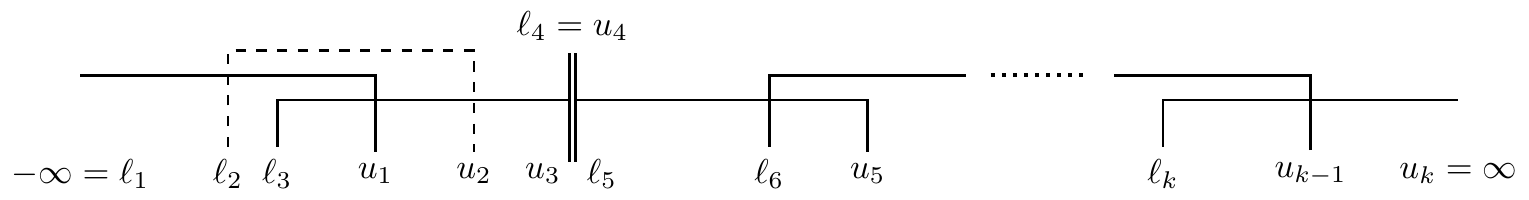}
	\caption{An example of an UNSAT covering.}\label{fig:example-covering}
\end{figure}

The input to Algorithm \ref{alg:construct-characterization} is an UNSAT covering $\I$ whose elements define intervals $I$ which together cover $\R$.  For an example of such a covering see \cref{fig:example-covering}.  There may be some redundancy here, like the second interval (from $\ell_2$ to $u_2$) in \cref{fig:example-covering} which is completely covered by the first and third intervals already.  To ensure soundness of our approach we need to remove at least those intervals which are included within a single other interval.  We discuss more details on dealing with redundancies in coverings in \cref{ssec:redundancy}.  

For now we simply make the reasonable assumption of the existence of an algorithm %\ref{alg:compute-cover},
\texttt{compute$\_$cover} which computes such a \emph{good covering} of the real line as a subset of an existing covering $\I$.  Since it is not crucial we will not specify the algorithm here, but we note that the ideas in \cite{JDF15} may be useful.

We assume that \texttt{compute$\_$cover} orders the intervals in its output according to the following total ordering:
\begin{equation}\label{order1}
	(\ell_1, u_1) \leq (\ell_2, u_2) \qquad \Leftrightarrow \qquad \ell_1 \leq \ell_2 \land (\ell_1<\ell_2 \lor u_1 \leq u_2)\ ,
\end{equation}
i.e. ordered on $\ell_i$ with ties broken by $u_i$.  
We will always have $\ell_1 = -\infty$ and $u_k = \infty$ with the remaining bounds defined as algebraic numbers (possibly not rational).
We further require that
\begin{equation}\label{order2}
(\ell_1, u_1) \leq (\ell_2, u_2) \qquad \Leftrightarrow \qquad \ell_1 \leq \ell_2 \land u_1 \leq u_2\ ,
\end{equation}
only possible if we exclude the cases where one interval is a subset of another.  Note that this is not just an optimisation but is actually crucial for the correctness of this approach, as explained in Section \ref{ssec:redundancy}.

The intervals we consider are either open (if $\ell \neq u$) or point intervals (if $\ell = u$). Note that it might make sense to extend the presented algorithm to allow for closed (or half-open) intervals as well, for example when built from weak inequalities. This could help to avoid some work on individual sample points like $\ell_4$ in \cref{fig:example-covering} and the point intervals we deal with in the examples in \cref{SEC:Example}. Such changes are straight-forward to implement and so we do not discuss them here to simplify the presentation.

\subsubsection{Constructing the characterisation}
\label{SSEC:Characterisation}

The first line of Algorithm \ref{alg:construct-characterization} ensures the ordering specified above, while the remainder uses CAD projection ideas to collect everything we need to ensure that the UNSAT covering $\I$ stays valid when we generalise the underlying sample point $s_i$ later on.
We include polynomials for a variety of reasons.
\begin{itemize}
\item First in \cref{proj-earlier} we pass along any polynomials with a lower main variable that had already been stored in the data structure.  These were essentially produced by the following steps at any earlier recursion, by an act of projection that skipped a dimension.  For example, the projection of any polynomial $f(x_1, x_3)$ into $(x_1,x_2)$-space will actually give a univariate polynomial in $x_1$.

\item Next in Lines \ref{proj-existence1} and \ref{proj-existence2} we identify polynomials that will ensure the existence of the current lower and upper bounds.
We add discriminants, whose zeros indicate where the original polynomial has multiple roots, and leading coefficients (with respect to the main variable), whose zeros indicate asymptotes of the original polynomial.  We may also require additional coefficients, as discussed in Section \ref{SSEC:Correctness}.   
If we ensure these polynomials do not change sign then we know that the algebraic varieties that defined $\ell$ and $u$ continue to exist (and no other varieties are spawned).

\item In Lines \ref{proj-overtakel} and \ref{proj-overtakeu} we generate polynomials whose sign-invariance ensures that $\ell$ and $u$ stay the \emph{closest} bounds.  I.e. we avoid the situation where they are undercut by those coming from some other variety.  We need only concern ourselves with those coming from the direction of the bound.  For example, when protecting an upper bound we need only worry about roots that are above it and take resultants accordingly on \cref{proj-overtakeu}.  This is because any root coming from below would need to first pass through the lower bound and the resultant from \cref{proj-overtakel} would block generalisation past that point.

\item In Lines \ref{proj-overlap1}$-$\ref{proj-overlap2} we finally derive resultants to ensure that the overlapping lower and upper bounds of adjacent intervals do not cross, which would \emph{disconnect} the UNSAT covering (leaving it not covering some portion of the line).  The correctness of this step requires an ordering with lack of redundancy, as specified above and discussed in detail in Section \ref{ssec:redundancy}.  This step also has the effect of ensuring the intervals do not overlap further to the extent that one then becomes redundant.
\end{itemize}

We note that the projection polynomials we collect are a subset of those collected by the McCallum projection operator for a full CAD \cite{McCallum1998}.  That operator would take the leading coefficient and discriminant of every polynomial involved, and all of the possible cross resultants.  Here we take only those relevant to the reasons for unsatisfiability.  Note that we have not taken the resultant of the polynomials that define the lower bound of an interval with those defining the upper bound of the same interval.  We explain why these are not necessary in Section \ref{SSEC:BoundsOfSameInterval} after discussing in detail the issue of interval redundancy.

Recall that we may have satisfiability over $s$ refuted by a single constraint in Algorithm \ref{alg:queryC} (i.e. the defining polynomial cannot change sign over $s$). In that case, after \cref{computercover} of \cref{alg:construct-characterization} the data structure $\I$ contains a single interval $(-\infty,\infty)$ %whose sets $L$ and $U$ only contain a single polynomial each
and we would have no resultants to compute.  %This matches our mathematical understanding that the discriminant and leading coefficient characterizes when such a constraint may cross zero around $s$.

\begin{algorithm}[t]
	\SetKwInOut{Input}{Input}\SetKwInOut{Output}{Output}
	\Input{Sample point $s \in \R^i$ and data structure $\I$ describing UNSAT covering over $s$ in dim.~$i{+}1$.}
	\Output{A set of polynomials $R \subseteq \mathbb{Q}[x_1,\dots,x_i]$ that characterizes a region around $s$ that is already unsatisfiable for the same reasons.}
	$\I :=$ \texttt{compute$\_$cover($\I$)} \nllabel{computercover} \tcp*{See \cref{SSEC:Ordering}}%\ref{alg:compute-cover}}
	%\tcp{Note that we now have $\I = [I_1, \dots, I_k]$ in ascending order of lower bounds}
	$R := \emptyset$ \;
	\ForEach{$I \in \I$}{
	    Extract $\ell=I_{\ell}, u=I_u, L=I_L, U=I_U, P_{i{+}1}=I_{P_{i{+}1}}, P_\bot=I_{P_\bot}$\;
	    $R := R \cup P_\bot$  \nllabel{proj-earlier}\;
		$R := R \cup \texttt{disc}(P_{i{+}1})$ \nllabel{proj-existence1}\;
		%$R := R \cup \{ \texttt{lcoeff}(p), \texttt{tcoeff}(p) \mid p \in P_{i{+}1} \}$ \nllabel{proj-existence2}\;	
		$R := R \cup \{\texttt{required}\_\texttt{coefficients}(p) \mid p \in P_{i{+}1} \}$ \nllabel{proj-existence2}\;			
		$R := R \cup \{ \texttt{res}(p,q) \mid p \in L, q \in P_{i{+}1}, q(s \times \alpha) = 0 \textrm{ for some } \alpha \leq l \}$ \nllabel{proj-overtakel}\;
		$R := R \cup \{ \texttt{res}(p,q) \mid p \in U, q \in P_{i{+}1}, q(s \times \alpha) = 0 \textrm{ for some } \alpha \geq u \}$ \nllabel{proj-overtakeu} \;
	}
	\For{$j \in \{1, \dots, |\I|-1\}$\nllabel{proj-overlap1}}{
		$R := R \cup \{ \texttt{res}(p,q) \mid p \in U_j, q \in L_{j{+}1} \}$ \nllabel{proj-overlap2}\;
	}
	Perform standard CAD simplifications to $R$ \label{standardSimp}\;
	\Return{$R$}
	
	\caption{\texttt{construct$\_$characterization($s, \I$)} \label{alg:construct-characterization}}
\end{algorithm}

\subsubsection{Simplification and bases of polynomials}
\label{SSEC:PolynomialBases}

Algorithm \ref{alg:construct-characterization} finishes in \cref{standardSimp} with some standard CAD simplifications to the polynomials.  These all stem from the fact that polynomials matter only so much as where they vanish.  E.g.:
\begin{itemize}
\item Remove any constants, or other polynomials than can easily be concluded to never equal zero.
\item Normalise the remaining polynomials to avoid storing multiple polynomials which define the same varieties.  E.g. multiply each polynomial by the constant needed to make it monic (i.e. divide by leading coefficient so for example polynomial $2x-1$ becomes $x-\tfrac{1}{2}$).  Other normal forms include the primitive positive integral with respect to the main variable.
\item Store a square free basis for the factors rather than the polynomials themselves (this much is necessary, else later resultants/discriminants will vanish); or even fully factorising (optional, but generally favoured for the efficiency gains it can bring).
\end{itemize}
We note that the original constraints are stored as presented for their analysis by Algorithm \ref{alg:queryC} but that when we store their defining polynomials in $\I$, we are actually storing the simplified bases of these polynomials described here.  Thus line 19 in Algorithm \ref{alg:queryC} from earlier is actually simplification as well as storage.

\subsubsection{Constructing the generalisation}
\label{SSEC:Generalisation}

Now let us discuss how this characterisation (set of polynomials) is used to expand the sample to an interval by Algorithm \ref{alg:interval-from-characterization}.  
We first separate the polynomials into $P_i$ and $P_\bot$ where $P_i$ are those polynomials that contain $x_i$ and $P_\bot$ the rest.  We then identify the crucial points over $s$ beyond which our covering may cease to be.  This step essentially evaluates the polynomials with main variable $x_i$ over the sample in $\mathbb{R}^{i{-}1}$ and calculates real roots of the resulting univariate polynomial.  There is some additional work within this sub-algorithm call on \cref{realroots} which we discuss in Section \ref{SSEC:Correctness}.

We then construct the interval around $s_i$ from the closest roots $\ell$ and $u$ and collect the polynomials that vanish in $\ell$ and $u$ in the sets $L$ and $U$, respectively.  We supplemented the real roots with $\pm \infty$ to ensure that $\ell$ and $u$ always exist, i.e. we can have $(-\infty,u)$ or $(\ell,\infty)$.  In this case the corresponding set $L$ or $U$ is empty.  % and some parts of the projection that follow will simplify accordingly.  

\begin{algorithm}[ht]
	\SetKwInOut{Input}{Input}\SetKwInOut{Output}{Output}
	\Input{Sample point $s \in \R^{i-1}$; an extension $s_i$ to the sample; and set of polynomials $P$ in $\mathbb{Q}[x_1,\dots,x_i]$ that characterize why $s \times s_i$ cannot be extended to a satisfying witness.}
	\Output{An interval $I$ around $s_i$ such that on $s \times I$ the constraints are unsatisfiable for the same reasons as on $s \times s_i$.}
	%where $I$ can also be $(-\infty,\infty)$.}
	$P_\bot := \{ p \in P \mid p \in \Q[x_1, ..., x_{i-1}] \}$ \;
	$P_i := P \setminus P_\bot$ \;
	%$Z :=$ \texttt{real$\_$roots}$(P_i, s)$ \nllabel{realroots}\;
%	\If{$Z = \emptyset$}{
%		\Return{$(-\infty,\infty,\emptyset,\emptyset,P_i,P_\bot)$}
		%\Return{$(\mathrm{FULL}, P_\bot \cup \texttt{disc}(P_i) \cup \texttt{lcoeff}(P_i))$}
%	}
	$Z := \{-\infty\} \cup \textrm{\texttt{real$\_$roots$\_$with$\_$check($P_i, s$)}} \cup \{\infty\}$ \nllabel{realroots}\;
	$l := \max \{ z \in Z \mid z \leq s_i \}$ \;
	$u := \min \{ z \in Z \mid z \geq s_i \}$ \;
	$L := \{ p \in P_i \mid p(s \times l) = 0 \}$ \;
	$U := \{ p \in P_i \mid p(s \times u) = 0 \}$ \;
	Define new interval $I$ with $I_\ell = \ell, I_u = u, I_L=L, I_U=U, I_{P_i}=P_i, I_{P_\bot}=P_\bot$\;
	\Return{I}
	
	\caption{\texttt{interval$\_$from$\_$characterization($s, s_i, P$)} \label{alg:interval-from-characterization}}
\end{algorithm}

In the case where $P_i$ has no real roots at all over $s$ then the UNSAT covering is valid unconditionally over $s$, and the interval $(-\infty, \infty)$ is formed and passed back to the main algorithm (where it could be taken to form the next covering in its entirety).

Let us briefly consider some simple examples for Algorithm \ref{alg:interval-from-characterization}.  Suppose we have variable ordering $x \prec y$, the partial sample $(x \mapsto 0)$ and that $P$ contains only the polynomial whose graph defines the unit circle: $x^2+y^2-1$.  Then \cref{realroots} forms the set $\{-\infty, -1, 1, \infty\}$.  
If $s_i$ had been the extension $(y \mapsto 0)$ then we would select $\ell = -1, u=1$, i.e. generalise to the $y$-axis inside the circle.  Similarly, if the extension had been $(y \mapsto 2)$ we would select $\ell=1, u = \infty$, i.e. generalise to the whole $y$ axis above the circle.  Finally, consider what would happen if $s_i$ had been the extension $(y \mapsto 1)$.  In that case, since $s_i$ is one of the roots in $Z$ it would be selected for both $\ell$ and $u$.  I.e. in that case no generalisation of $s_i$ is possible.

\subsubsection{Correctness of the generalisation}
\label{SSEC:Correctness}

The correctness of Algorithms \ref{alg:construct-characterization} and \ref{alg:interval-from-characterization} relies on CAD theory via McCallum projection, with Algorithm \ref{alg:construct-characterization} an analogue of projection and Algorithm \ref{alg:interval-from-characterization} an analogue of lifting.  However, there are a number of subtleties.

First, regarding exactly which coefficients are computed in Algorithm \ref{alg:construct-characterization} \cref{proj-existence2}.  As explained above, the leading coefficient is essential to include as its vanishing would indicate an asymptote.  However, we must also consider what happens within regions of such vanishing: there the polynomial has reduced in degree, and so a lesser coefficient is now leading and must be recorded for similar reasons.  We should take successive coefficients until one is taken that can be easily concluded not to vanish (e.g. it is constant) or if it may be concluded that the included coefficients cannot vanish simultaneously.  In our context this is simpler: so long as a coefficient evaluates at the sample point to something non-zero then we need not include any subsequent coefficients (since by including that coefficient in the characterisation we ensure we do not generalise to where it is zero).   We do need to include preceding coefficients that were zero as we must ensure that the generalisation does not cause them to reappear. This is described by Algorithm \ref{alg:reqCoeff}.

\begin{algorithm}[ht]
	\SetKwInOut{Input}{Input}\SetKwInOut{Output}{Output}
	\Input{Sample point $s \in \R^i$ and set of polynomials $P_{i+1}$ each with main variable $x_{i+1}$.}
	\Output{A set of polynomials in $\mathbb{Q}[x_1,\dots,x_i]$ consisting of those coefficients of $p$ required for the characterization around $s$.}
    $C = \emptyset$\;	
	\ForEach{$p \in P_{i+1}$}{
	    \While{$p \neq 0$}{
	    $lc := \textrm{\texttt{lcoeff}}(p)$\;
	    $C  := C \cup \{lc\}$\;
	    \If{$lc$ evaluated at $s$ is non zero}{
	    	\textbf{Break} the while loop\;
	    }	 
	    $p := p - \textrm{\texttt{lterm}}(p)$\;
	    }
	}
	\Return{$C$}
	\caption{\texttt{required$\_$coefficients($s, P_{i+1}$)} \label{alg:reqCoeff}}
\end{algorithm}

With such calculation of coefficients our characterisations are then subsets of the McCallum projection \cite{McCallum1998}: we ensure that individual polynomials are delineable but we cannot claim that for the characterisation as a set since we differ from \cite{McCallum1998} in which resultants are computed.  The inclusion of resultants in CAD projection is to ensure a constant structure of intersections between varieties in each cell.  McCallum projection takes all possible resultants between polynomials involved.  Instead, we take only those needed to maintain the structure of our covering, as detailed by the arguments in Section \ref{SSEC:Characterisation}. 

\subsubsection{Completeness of the generalisation}
\label{SSEC:Completeness}

McCallum projection is not \emph{complete}, i.e. there are some (statistically rare) cases where its use is known to be invalid, and these could be inherited in our algorithm.  The problem can occur when a polynomial vanishes at a sample of lower dimension (known as \emph{nullification}) potentially losing critical information.  For example, consider a polynomial $zy-x$ which vanishes at the lower dimensional sample $(x,y) = (0,0)$.  The polynomials' behaviour clearly changes with $z$ but that information would be lost.

Such nullifications can be easily identified when they occur.  We assume that the sub-algorithm used in Algorithm \ref{alg:interval-from-characterization} \cref{realroots} will inform the user of such nullification.  What should be done in this case?  An extreme option would be to recompute the characterisation to include entire subresultant sequences as in Collins' projection \cite{Collins1975}, or to use the operator of Hong \cite{Hong1990}.  A recent breakthrough in CAD theory could offer a better option:  \cite{MPP19} proved that Lazard projection \cite{Lazard1994} is complete\footnote{The proofs in Lazard's original 1994 paper were found to be flawed and so the safe use of this operator comes only with the 2019 work of McCallum, Parusi\'{n}ski and Paunescu \cite{MPP19}.}.  The Lazard operator includes leading and trailing coefficients, and requires more nuanced lifting computations. Instead of evaluating polynomials at a sample and calculating real roots of the resulting univariate polynomials, we must instead perform a Lazard evaluation \cite{MPP19} of the polynomial at the point.  This will substitute the sample coordinate by coordinate, and in the event of nullification divide out the vanishing factor to allow the substitution to continue.  Thus no roots are lost through nullification.  

We have not yet adapted our algorithm to Lazard theory:  although SMT-RAT already has an implementation of Lazard evaluation, we are not yet clear on how the polynomial identification in Algorithm \ref{alg:interval-from-characterization} should be adapted in cases of nullification.  Also, it is not trivial to argue the safe exclusion of trailing coefficients in cases where the leading coefficient is constant, as it is with McCallum, since the underlying Lazard delineability relies heavily on properties of the trailing coefficient.  Our current implementation is hence technically incomplete, however, we can produce warnings for such cases.  The experiments on the SMT-LIB detailed in Section \ref{SEC:Experiments} show that these nullifications are a very rare occurrence. 

% JHD to work out what needs doing to l/u/L/U in the event we do Lazard.

\subsection{Redundancy of Intervals}
\label{ssec:redundancy}

\begin{figure}[ht]
  \begin{center}
    \begin{subfigure}[b]{0.49\textwidth}
      \centering
      \includegraphics{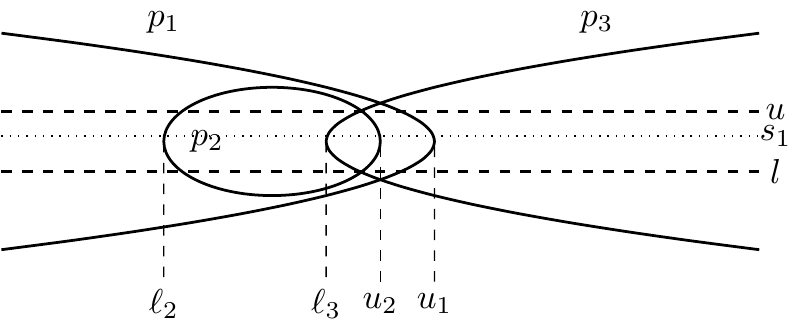}
    \caption{Redundancy that still yields resultants.}\label{fig:redundant-intervals-first-kind:1}
  \end{subfigure}
		\begin{subfigure}[b]{0.49\textwidth}
		  \centering
                            \includegraphics{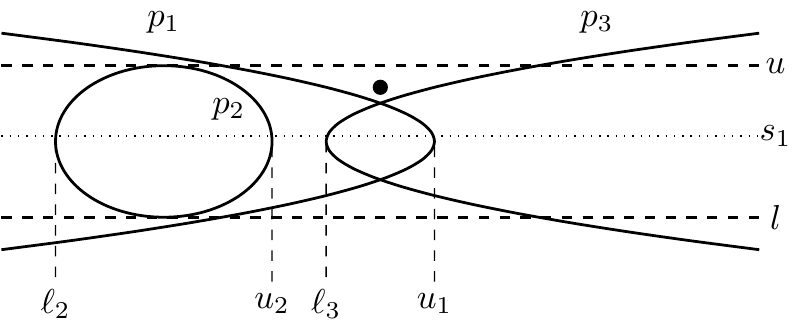}
			\caption{Redundancy that yields no resultants.}\label{fig:redundant-intervals-first-kind:2}
		\end{subfigure}
	\end{center}
	\caption{Possible situations for redundant intervals of the first kind.\label{fig:redundant-intervals-first-kind}}
\end{figure}

\begin{figure}[ht]
	\begin{center}
		\begin{subfigure}[b]{0.49\textwidth}
		  \centering
                  \includegraphics{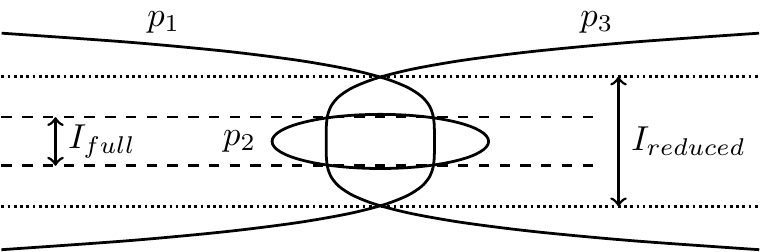}
			\caption{Beneficial reduction.}\label{fig:redundant-intervals-second-kind:1}
		\end{subfigure}
		\begin{subfigure}[b]{0.49\textwidth}
		  \centering
                  \includegraphics{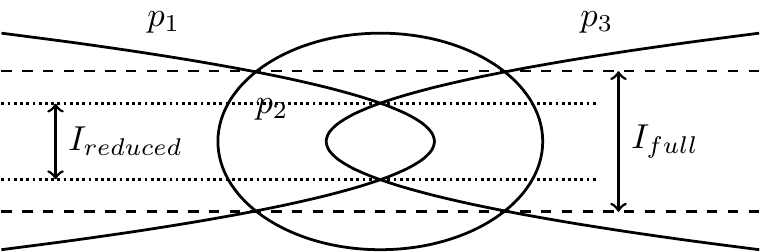}
			\caption{Detrimental reduction.}\label{fig:redundant-intervals-second-kind:2}
		\end{subfigure}
	\end{center}
	\caption{Possible situations for redundant intervals of the second kind.\label{fig:redundant-intervals-second-kind}}
\end{figure}

We discussed in \cref{SSEC:Ordering} that the set of intervals in a covering may contain redundancies.  We distinguish between two possibilities for how an interval $I$ can be redundant:
\begin{enumerate}
\item $I$ is covered by a single other interval entirely, as interval $(\ell_2, u_2)$ is by interval $(\ell_1, u_1)$ in \cref{fig:redundant-intervals-first-kind}.
\item $I$ is covered through multiple other intervals, as the interval defined by the bounds of $p_2$ would be by those from $p_1$ and $p_3$ in \cref{fig:redundant-intervals-second-kind}. 
\end{enumerate}
We now explain why we need to remove redundancies of the first kind, but the second could be kept.

\subsubsection{Redundancy of the first kind}

%Let us first consider the case where an interval is covered by a single other interval as shown in \cref{fig:redundant-intervals-first-kind}.
The resultants we produce in Lines \ref{proj-overlap1}$-$\ref{proj-overlap2} of \cref{alg:construct-characterization} are meant to ensure that consecutive intervals (within the ordering) continue to overlap on the whole region we exclude. Thus we must guarantee that they overlap on our sample point in the first place.

In the examples of \cref{fig:redundant-intervals-first-kind} we assume constraints which are satisfied only in the regions outside of the graphed curves, and so the point marked in \cref{fig:redundant-intervals-first-kind:2} is a satisfying witness.  The two examples differ only by a small change to the polynomial $p_2$: in \cref{fig:redundant-intervals-first-kind:1} $p_2$ intersects $p_3$ while in \cref{fig:redundant-intervals-first-kind:2} it does not.  For both examples the numbering of the intervals means we will calculate the resultants res($p_1, p_2$), res($p_2, p_3$) but not res($p_1, p_3$).  In each case the bounds obtained are indicated by the dashed lines: in  \cref{fig:redundant-intervals-first-kind:1} they come from the roots of  res($p_2, p_3$) while in \cref{fig:redundant-intervals-first-kind:2} this resultant has no real roots and the closest bounds instead come from the discriminant of $p_2$.

So, in \cref{fig:redundant-intervals-first-kind:1} the excluded region is all unsatisfiable, i.e. correct, but only by luck! The resultant that bounds the excluded regions has no relation to the actual bound (the intersection of $p_1$ and $p_3$).  In \cref{fig:redundant-intervals-first-kind:2} we exclude too much and erroneously exclude the dot which actually marks a satisfying sample.

We do not see an easy way to distinguish the two situations presented in \cref{fig:redundant-intervals-first-kind} and we therefore excluded all redundancies of this kind in \cref{SSEC:Ordering}, by requiring that we have the stronger ordering (\ref{order2}) rather than just the weaker version (\ref{order1}).

\subsubsection{Redundancy of the second kind}

The error described above of excluding a satisfiable point because of a redundancy of the first kind, could not occur in the presence of a redundant interval of the second kind.  The algorithm assumes that every adjacently numbered interval overlaps, which is not the case for a redundancy of the first kind but is the case for one of the second kind.  

However, should we still remove redundant intervals of the second kind for efficiency purposes?  Removal would mean less intervals in the covering, and thus less projection in Algorithm \ref{alg:construct-characterization} and less root isolation in Algorithm \ref{alg:interval-from-characterization}.  However, it does not mean the generalisation has to be bigger.  

Intuitively, reducing such a redundancy makes the overlap between adjacent intervals smaller. But this may also mean that the interval we can exclude in the lower dimension is smaller than it would have been if we kept the redundant interval, retaining a larger overlap.  Consider the two examples from \cref{fig:redundant-intervals-second-kind}: the polynomials differ but the geometric situation and position of intervals is similar.  In both cases an interval defined by the bounds of $p_2$ will be redundant when combined with those from the bounds of $p_1$ and $p_3$.  If we do not reduce the covering by the redundant interval then we exclude $I_{full}$ but if we do we exclude $I_{reduced}$. The excluded region would grow from reduction in \cref{fig:redundant-intervals-second-kind:1} but shrink in \cref{fig:redundant-intervals-second-kind:2}.

We do not see an easy way to check which situation we have (other than completely calculating both and taking the better one).  The decision of whether to reduce could be taken heuristically by an implementation.  Further investigation into this would be an interesting topic for future work.

\subsubsection{Upper and lower bounds of the same interval crossing}
\label{SSEC:BoundsOfSameInterval}

\begin{figure}
	\centering
        \includegraphics[width=0.6\textwidth]{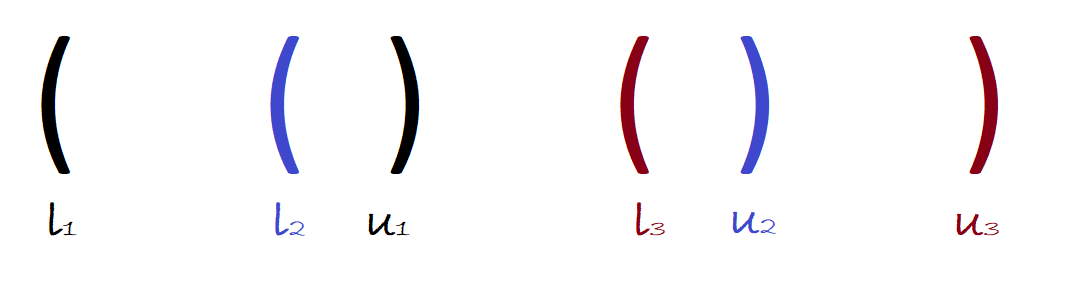}
        \includegraphics[width=0.6\textwidth]{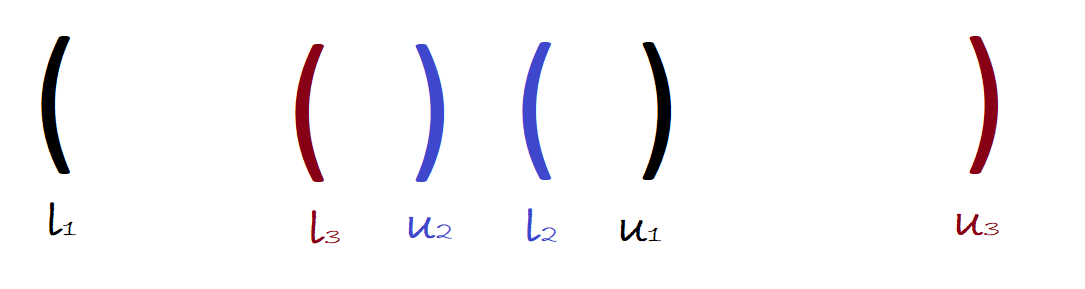}        
	\caption{An example to show that if an interval's upper and lower bounds cross then the interval is no longer required.}\label{fig:IntervalBoundsCrossingExample}
\end{figure}

Recall that at the end of Section \ref{SSEC:Characterisation} we noted that our characterisation does not include the resultant of polynomials defining the upper bound of an interval with those defining the lower bound of the same interval.  Now we have discussed redundancy we can explain why these are not required.   Consider for example the triple of intervals in the top of Figure \ref{fig:IntervalBoundsCrossingExample} and the possibility that $\ell_2$ and $u_2$ may swap order (which would have been blocked by taking the resultant of defining polynomials in the characterisation).  Now, since the characterisation did include the resultants of polynomials defining upper bounds with those defining lower bounds of the next interval it is not possible for $\ell_2$ to pass to the right of $u_1$, or for $u_2$ to pass to the left of $\ell_3$.  Thus the only way that $\ell_2$ and $u_2$ could pass in a generalisation is if their neighbours moved with them, as in the bottom of Figure \ref{fig:IntervalBoundsCrossingExample}. We can now observe that the second interval has become redundant, i.e. the portion of the line that it used to cover is now fully covered by the other two intervals.  The bounds for the second interval must now be fully contained by both the first \emph{and} second interval.  Thus at all times in this situation the first and third interval must overlap.  Hence this redundancy is of the second kind and thus safe for the correctness of the algorithm. 

\subsection{Embedding as Theory Solver}
\label{SUBSEC:TheorySolver}

Our algorithm can be used as a standalone solver for a set of real arithmetic constraints, but our main motivation is to use it as a theory solver within an SMT solver.  Such theory solvers should be \emph{SMT compliant}:
\begin{itemize}
\item They should work \emph{incrementally}, i.e allow for the user to add constraints and solve the resulting problem with minimal recomputation.
\item They should similarly allow for \emph{backtracking}, i.e. the removal of constraints by the user.
\item They should construct reasons for unsatisfiability, which usually refers to a subset of the constraints that are already unsatisfiable, often referred to as \emph{infeasible subsets}.
\end{itemize}

For infeasible subsets we store, for every interval, a set of constraints that contributed to it, in a similar way to what was called \emph{origins} in~\cite{Kremer2019}.  For intervals created in \cref{alg:queryC} we simply use the respective constraint $c$ as their origin.  For intervals that are constructed in \cref{alg:interval-from-characterization} the set of constraints is computed as the union of all origins used in the corresponding covering in \cref{alg:construct-characterization}.  The constraints are then gathered together as the final step before returning an UNSAT result in Algorithm \ref{alg:user} (Line \ref{line:unsatcore}).

We have yet to implement incrementality and backtracking, but neither pose a theoretical problem.  
Though possibly involved in the implementation, the core idea for incrementality is straight-forward: after we found a satisfying sample we retain the already computed (partial) coverings for every variable and extend them with more intervals from the newly added constraints.  
For backtracking, we need to remove intervals from these data structures based on the input constraints they stem from. As we already store these to construct infeasible subsets, this is merely a small addition.

\section{Worked Examples}
\label{SEC:Example}

\subsection{Simple SAT Example in 2D}
\label{SSEC:SimpleEx}

\begin{figure}[ht]
	\centering	
        \includegraphics{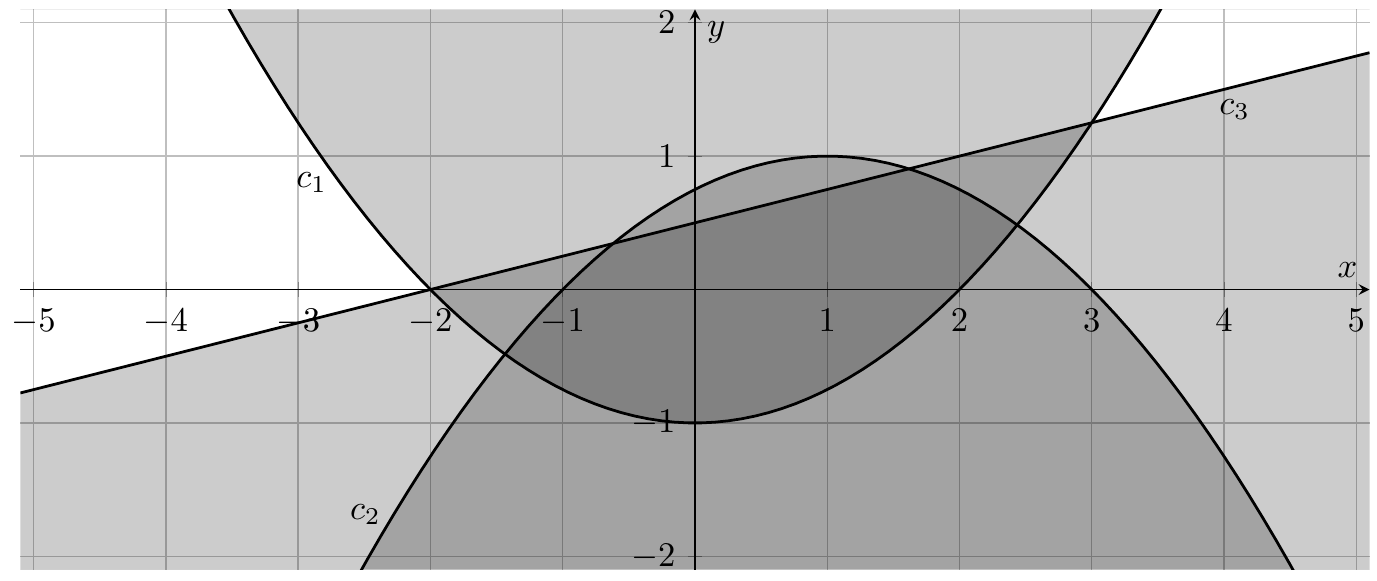}
	\caption{Graphs of the polynomials defining the three constraints in the simple example (\ref{eq:simple}).}\label{fig:example-simple}
\end{figure}

We start with a simple two-dimensional example to show how the algorithm proceeds in general.  Our aim is to determine the satisfiability for the conjunction of the following three constraints:  
\begin{align}
		c_1 \, : \quad 4 \cdot y < x^2- 4, 		\qquad 
		c_2 \, : \quad 4 \cdot y > 4 - (x-1)^2,  \qquad 
		c_3 \, : \quad 4 \cdot y > x + 2.
\label{eq:simple}
\end{align}
The three defining polynomials are graphed in \cref{fig:example-simple}, with the unsatisfiable regions for each adding a layer of shading (so the white regions are where all three constraints are satisfied).  

We now simulate how our algorithm would process these constraints, under variable ordering $x \prec y$, to find a point in one of those satisfying regions.  
The user procedure (Algorithm \ref{alg:user}) starts by calling the main algorithm \texttt{get\_unsat\_cover()} (Algorithm \ref{alg:main}) with an empty tuple as the sample.

\paragraph{\texttt{get\_unsat\_cover($s = ()$)}}
  Since all the constraints are bivariate, the call to \texttt{get\_unsat\_intervals()} cannot ascertain any intervals to rule out, initialising $\I = \emptyset$.  We thus enter the main loop of Algorithm \ref{alg:main} and sample $s_1 = 0$ in \cref{s2}.  This sample does not have  full dimension and we perform the recursive call in \cref{recurse}.

\paragraph{\texttt{get\_unsat\_cover($s = (x \mapsto 0)$)}}
This time, the call of \texttt{get\_unsat\_intervals($s = (x \mapsto 0)$)} sees all three constraints rendered univariate by the sample.  Each univariate polynomial has one real root and thus decomposes the real line into three intervals.  
\begin{center}
\begin{tabular}{lll}
Constraint & Real Roots & Intervals \\ 
$c_1 \, : \quad 4 \cdot y < x^2- 4$ 		
& $\{ -1 \}$ 	& $(-\infty,-1),\,		[-1,-1],\,		(-1, \infty)$ \\
$c_2 \, : \quad 4 \cdot y > 4 - (x-1)^2$ 	
& $\{ \tfrac{3}{4} \}$ 	& $(-\infty,\tfrac{3}{4}),\,	[\tfrac{3}{4},\tfrac{3}{4}],\,	(\tfrac{3}{4}, \infty)$ \\
$c_3 \, : \quad 4 \cdot y > x + 2$ 		
& $\{ \tfrac{1}{2} \}$ 	& $(-\infty,\tfrac{1}{2}),\,[\tfrac{1}{2},\tfrac{1}{2}],\,	(\tfrac{1}{2}, \infty)$ \\
\end{tabular}
\end{center}
After analysing each of these 9 regions we identify 6 for which a constraint is infeasible.  Thus $\I$ is initialised as a set of six intervals as below (where $p_i$ refers to the defining polynomial for constraint $c_i$).
%\begin{align*}
%	\I = \, \{  \quad
%	& (-1,		&& -1,		&& \{p_1\},&& \{p_1\},&& \{p_1\},&& \emptyset), & \\
%	& (-1,		&& \infty,	&& \{p_1\},&& \emptyset,&& \{p_1\},&& \emptyset), & \\
%	& (-\infty,	&& \tfrac{3}{4},	&& \emptyset,&& \{p_2\},&& \{p_2\},&& \emptyset), & \\
%	& (\tfrac{3}{4},	&& \tfrac{3}{4},	&& \{p_2\},&& \{p_2\},&& \{p_2\},&& \emptyset), & \\
%	& (-\infty,	&& \tfrac{1}{2},		&& \emptyset,&& \{p_3\},&& \{p_3\},&& \emptyset), & \\
%	& (\tfrac{1}{2},		&& \tfrac{1}{2},		&& \{p_3\},&& \{p_3\},&& \{p_3\},&& \emptyset)  & \}.
%\end{align*}
\begin{align*}
I_1: \quad & \ell = -1,		&& u=-1,		
&& L=\{p_1\},&& U=\{p_1\},
&& P_2 = \{p_1\},&& P_\bot=\emptyset, & 
\\
I_2: \quad & \ell=-1,		&& u=\infty,	
&& L=\{p_1\},&& U=\emptyset,
&& P_2=\{p_1\},&& P_\bot=\emptyset, & 
\\
I_3: \quad & \ell=-\infty,	&& u=\tfrac{3}{4},	
&& L=\emptyset,&& U=\{p_2\},
&& P_2 =\{p_2\},&& P_\bot=\emptyset, & 
\\
I_4: \quad & \ell=\tfrac{3}{4},	&& u=\tfrac{3}{4},	
&& L=\{p_2\},&& U=\{p_2\},
&& P_2 =\{p_2\},&& P_\bot=\emptyset, & 
\\
I_5: \quad & \ell=-\infty,	&& u=\tfrac{1}{2},		
&& L=\emptyset,&& U=\{p_3\},
&& P_2 =\{p_3\},&& P_\bot=\emptyset, & 
\\
I_6: \quad & \ell=\tfrac{1}{2},		&& u=\tfrac{1}{2},		
&& L=\{p_3\},&& U=\{p_3\},
&& P_2 =\{p_3\},&& P_\bot=\emptyset.  & 
\end{align*}
Although $(-\infty,\infty)$ is not an interval of $\I$ we observe that $\R$ is already covered by the two intervals $(-\infty,\tfrac{1}{2})$ and $(-1, \infty)$.  Note how the second of the three constraints is not part of the conflict which simplifies the following calculations.

Since the real line is already covered we do not enter the main loop of Algorithm \ref{alg:main} in this call.  Instead we immediately proceed to the final line where we return (UNSAT, $\I$) to the call of \texttt{get\_unsat\_cover($s = ()$)}.  Since the flag is UNSAT the next step in that call is constructing a characterisation in \cref{algCJ}. 

\paragraph{\texttt{construct\_characterisation($s = (x \mapsto 0), \I$)}}
As already observed, the intervals $(-\infty,\tfrac{1}{2})$ and $(-1, \infty)$ cover $\R$ and thus the call to \texttt{compute\_cover()} at the start should simplify $\I$ to the following:
\begin{align*}
I_1: \quad & \ell=-\infty,&& u=\tfrac{1}{2},
&& L=\emptyset,&& U=\{4 \cdot y - x -2\},
&& P_2=\{4 \cdot y - x -2\},&& P_\bot=\emptyset, 
\\
I_2: \quad & \ell=-1,&& u=\infty,
&& L=\{4 \cdot y - x^2 + 4\},&& U=\emptyset,
&& P_2=\{4 \cdot y - x^2 + 4\},&& P_\bot=\emptyset.
\end{align*}

As $P_i$ only contains a single polynomial in each interval there are no resultants to calculate in the first loop, and only a single one from the second:
\[
\texttt{res}_y(4 \cdot y - x -2, 4 \cdot y - x^2 + 4) = -4\cdot x^2 + 4 \cdot x + 24.
\]
Further, $P_\bot = \emptyset$ and the discriminants and leading coefficients all evaluate to constants:
\begin{align*}
	& \texttt{disc}_y(4 \cdot y - x -2) = 1,    && \texttt{lcoeff}_y(4 \cdot y - x -2) = 4, \\
	& \texttt{disc}_y(4 \cdot y - x^2 + 4) = 1, && \texttt{lcoeff}_y(4 \cdot y - x^2 + 4) = 4.
\end{align*}
So the output set from \texttt{construct\_characterisation($s = (x \mapsto 0), \I$)} consists of a single polynomial:
$R = \{ x^2 - x - 6 \}$.
Then in the original call to \texttt{get\_unsat\_cover($s = ()$)} we continue working in \cref{algIFJ} with the construction of an interval from the characterisation.

\paragraph{\texttt{interval\_from\_characterisation($s = (), s_i = (x \mapsto 0), P = \{ x^2 - x -6 \}$)}}
We see that $P_\bot = \emptyset$ and $P_i = P$ and obtain the real roots $\{ -2, 3 \}$. Consequently we obtain $l = -2$, $u = 3$ with both $L$ and $U$ as $P_i$.  We thus return the unsatisfiable interval $(-2, 3, P_i, P_i, P_i, \emptyset)$.

Back in our initial call to Algorithm \ref{alg:main} we add this interval to $\I$ and continue with a second iteration of the main loop.  This time we must sample some value of $x$ outside of $(-2, 3)$, for example $x=-3$ or $x=4$, both of which can be extended to a full satisfying sample: $(x \mapsto -3, y \mapsto 0)$ and $(x \mapsto -4, y \mapsto 2)$ respectively. In fact, any sample for $x$ outside of the interval $(-2, 3)$ can be extended to a full assignment and that would be always discovered during the next recursive call to Algorithm \ref{alg:main}.

\paragraph{\texttt{get\_unsat\_cover($s = (x \mapsto \hat{x})\}$)}}  Here $\hat{x}$ is the sample for $x$ outside $(-2,3)$ chosen above.  For any such sample first Algorithm \ref{alg:interval-from-characterization} will rule out any extension for $y$ that is infeasible and then \texttt{sample\_outside} would pick a satisfying value of $y$ in the first iteration of the main loop.  This would form a full dimensional sample which is returned along with the flag SAT by \cref{retSAT1}.

Back in our initial call to Algorithm \ref{alg:main} we then pass the tuple of flag and satisfying sample back to the user function in the return on \cref{retSAT2}.

\subsubsection{Comparing to Incremental CAD}

Let us consider how this would compare with the incremental version of traditional CAD.  Recall that this performs one projection step at a time, and then refines the decomposition with respect to the output (producing extra samples).  Thus the computation path depends on the order in which projections are performed.  The discriminants and coefficients of the input do not contribute anything meaningful to the projections, so it all depends in which order the resultants are computed.  If the implementation were unlucky and picked the least helpful resultant it will end up performing more decomposition than is required.  For this particular example the effect is mitigated a little by the implementation's preference for picking integer sample points allowing it to find a satisfying sample earlier than guaranteed by the theory:  i.e. the cell being formed by the decomposition is not truth invariant for all constraints, but by luck the sample picked is SAT and so the algorithm can terminate early.  So for this example our incremental CAD performs no more work than the new algorithm, but that is through luck rather than guidance.  In contrast, the superiority of the new algorithm over incremental CAD is certain for the next example.   

\subsubsection{Comparison to NLSAT}

We may also compare to the NLSAT method~\cite{JovanovicdeMoura2012a} which also seeks a single satisfying sample point for all the constraints.  Like our new method, NLSAT is model driven starting with a partial sample; it \emph{explains} inconsistent models locally using CAD techniques; and thus exploits the locality to exclude larger regions; combining multiple of these explanations to determine unsatisfiability.  The difference is that the conflicts are then converted into a new lemma for the Boolean skeleton and passed to a separate SAT solver to generate a new model.  
The implementation of NLSAT in SMT-RAT (see Section \ref{SSEC:OtherSolvers}) performs the following steps for the first example:
\begin{align*}
	\textrm{theory model} \quad & \textrm{explanation clause} & \textrm{excluded interval} \\
	x \mapsto 0 \quad & (c_1 \land c_2) \rightarrow (x \leq \underline{-1.44} \lor \underline{2.44} \leq x) & (\underline{-1.44},\underline{2.44}) \\
	x \mapsto -2 \quad & (c_1 \land c_3) \rightarrow (x \neq -2) & [-2,-2] \\
	x \mapsto 3 \quad & (c_1 \land c_3) \rightarrow (x \neq 3) & [3, 3] \\
	x \mapsto -3 \quad & y \mapsto 0
\end{align*}

The boundaries of the first excluded interval are the two real roots of the polynomial $2x^2 - 2x - 7$ which is the resultant of the defining polynomials for $c_1$ and $c_2$.  Rather than give these as surds or algebraic numbers we use the decimal approximation simply to shorten the presentation in the paper.  Throughout, a decimal underlined refers to a full algebraic number that we have computed but choose not to display for brevity.

For this example NLSAT took three conflicts to find a satisfying model (compared to two for our new algorithm).  However, the difference is due to luck.  In the first iteration NLSAT was unlucky to select a subset of constraints that rules out a smaller UNSAT region. It could have instead chosen $c_1$ and $c_3$ in the first iteration, essentially yielding the very same computation as our method. Conversely our algorithm could also have chosen $c_1$ and $c_2$ as a cover and then have needed another iteration. 

\subsubsection{Comparison to NuCAD}

\begin{figure}[ht]
	\begin{center}
          \includegraphics{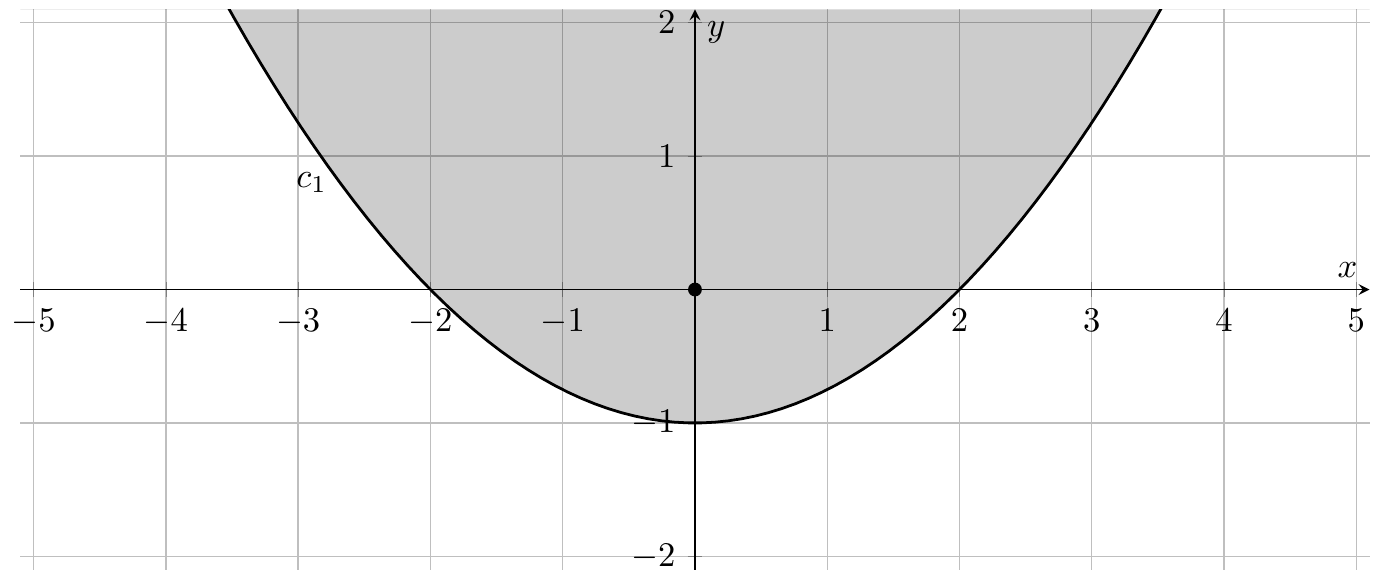}
	\end{center}

	\caption{First NuCAD cell for $(0,0)$.}\label{fig:example-simple-nucad:1}
\end{figure}

Our algorithm as stated is designed to solve satisfiability problems, and thus terminates as soon as a satisfying sample is found.  Thus it does not compare directly with NuCAD which builds an entire decomposition of $\R^n$ relative to the truth of a quantifier free logical formula which can then be used to solve more general QE problems stated about that formula\footnote{Although this does require some additional computation as outlined in \cite{Brown2017}.}.  
NuCAD constructs the decomposition of one cell (with known truth-value) in $\R^n$ at a time, so there is a natural variant of the algorithm where we construct cells only until we find one in which the input formula is \tru.  Consider how NuCAD might proceed for the simple example.  
A natural starting point would be to consider the origin first. NuCAD would recognise that constraints are not satisfied here and choose one of them to process.  Let us assume NuCAD chooses in the order the constraints are labelled (i.e. pick $c_1$); then like us it would generalise this knowledge beyond the sample and decompose the plane as in \cref{fig:example-simple-nucad:1}.  The shaded cell is known to be UNSAT while the white region is a cell whose truth value is as yet unknown.

\begin{figure}[ht]
	\centering
        \includegraphics{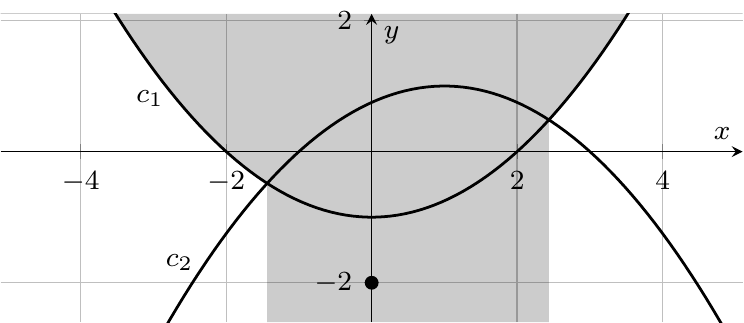}
        $\quad$
        \includegraphics{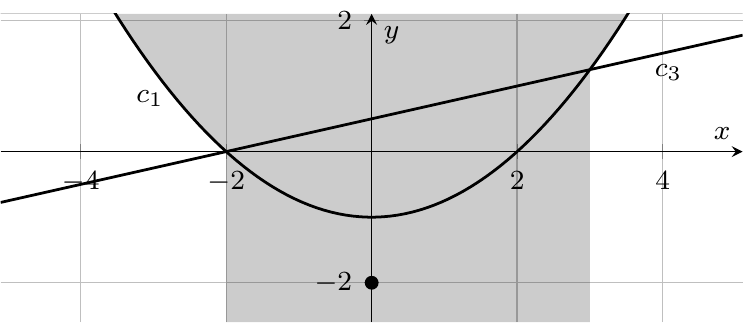}
	\caption{Possible continuations for NuCAD after \cref{fig:example-simple-nucad:1}.}\label{fig:example-simple-nucad:2}
\end{figure}

Now NuCAD must pick a new sample outside of the shaded cell.  A natural choice would be keep one coordinate zero (i.e. move along the axis).  If it were to move left along the $x$ axis and pick the first integer outside the shaded cell, i.e. sample $(-3,0)$ then it would find a satisfying point and terminate, but if it were to move right to $(3,0)$ it would have to decompose further.  Another natural strategy may be to keep the $x$ value fixed and try other $y$ values, which allows a direct comparison to our method.  A preference for the next integer leads to the sample $(0,-2)$ where both $c_2$ and $c_3$ are violated.  The two possible resulting cells are shown in \cref{fig:example-simple-nucad:2} where picking $c_2$ leads to the decomposition on the left, and picking $c_3$ to the decomposition on the right (similar to the two possible steps our algorithm had).    

It is possible under a natural strategy for NuCAD to find a satisfying point on its second sample, but this is not guaranteed by the theory.  In comparison, our algorithm is more guided, if starting at the origin then it could not take more than three iterations for this example.

\subsection{More Involved UNSAT Example in 2D}\label{sec:second-example}
\label{SSEC:SecondEx}

In the first example above we saw that the new algorithm was able to learn from conflicts to guide itself towards a satisfying sample.  However, due to luck and sensible implementation choices the alternatives could process the example with a similar amount of work to the new algorithm.  So we next present a more involved example that will make clearer the savings offered by the new approach.  This example will ultimately turn out to be unsatisfiable and thus the incremental CAD of SMT-RAT will in the end perform all projection and produce a regular CAD, as for example would be produced by a traditional CAD implementation like QEPCAD.  The advantage of our algorithm in this case is that we can avoid some parts of the projection that are the most expensive.  

\begin{figure}[ht]
	\centering
        \includegraphics{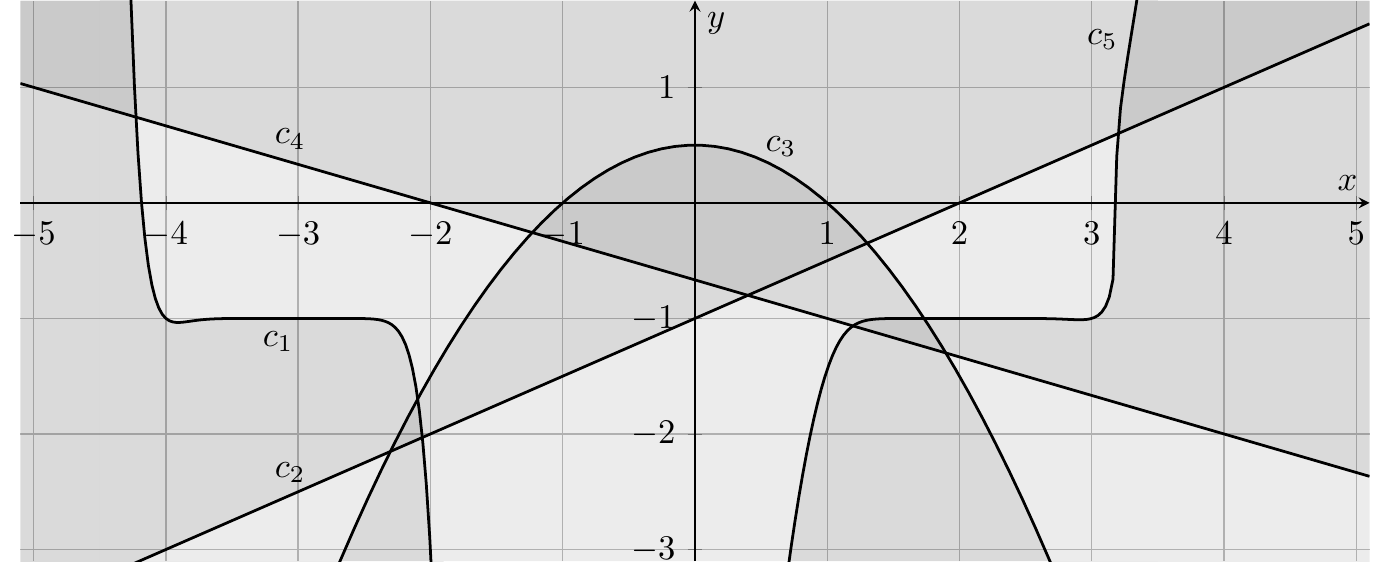}
	\caption{Graphs of the polynomials defining the constraints in the more involved example (\ref{eq:second-example}).\label{fig:example-compare-to-cad}
	}
\end{figure}

Our aim for this example is to determine the satisfiability for the conjunction of the following five constraints:
\begin{align}
	c_1&: y > (-x-3)^{11} - (-x-3)^{10} - 1 &
	c_2&: 2 \cdot y < x - 2 &
	c_3&: 2 \cdot y > 1 - x^2 \nonumber \\
	c_4&: 3 \cdot y < -x - 2 &
	c_5&: y^3 > (x-2)^{11} - (x-2)^{10} - 1 
\label{eq:second-example}
\end{align}
They are depicted graphically in \cref{fig:example-compare-to-cad}, with each constraint again adding a layer of shading where it is unsatisfiable.  This time we see there is no white space and so together the constraints are unsatisfiable.

Before we proceed let us remark on how the example was constructed.  There are two constraints of high degree ($c_1$ and $c_5$) while the rest are fairly simple.  The problem structure was chosen so that $c_1$ and $c_2$ conflict on the left part of the figure, $c_2$ and $c_3$ in the middle and $c_4$, and $c_5$ on the right. Thus while both $c_1$ and $c_5$ are important to determining unsatisfiability, they never need to be considered together to do this.  I.e. we have no need of their resultant (with respect to $y$):  an irreducible degree 33 polynomial in $x$ which is dense (34 non-zero terms) with coefficients that are mostly 20 digit integers.    
In this example $c_1$ and $c_5$ are of degree $11$ but we could generalise the example by increasing the degree of these polynomials arbitrarily while retaining the underlying problem structure which allows us to avoid working with them together.  Hence the advantage over algorithms which do consider them together can be made arbitrarily large.

We now describe how our new algorithm would proceed for this example but skip some details compared to the explanation of the previous example, to avoid unnecessary repetition.  %Real roots in this description are provided subject to rounding as their actual value is irrelevant here (all constraints are strict inequalities)  but note that the algorithm can compute with full algebraic numbers when needed.

\paragraph{\texttt{get\_unsat\_cover($s = ()$)}}
As all constraints contain $y$ we get no unsatisfiable intervals in the first dimension. We enter the loop and sample $s_1 = 0$ and enter the recursive call.

\paragraph{\texttt{get\_unsat\_cover($s = (x \mapsto 0)$)}}
All constraints are univariate and \texttt{get\_unsat\_intervals($s = (x \mapsto 0)$)} returns $\I$ defining the following $10$ unsatisfiable intervals:
\begin{align*}
	& (-\infty, -236197),	&& [-236197, -236197], 	&& \textrm{from } c_1, \\
	& [-1, -1],				&& (-1, \infty), 		&& \textrm{from } c_2, \\
	& (-\infty, \tfrac{1}{2}),		&& [\tfrac{1}{2}, \tfrac{1}{2}], 			&& \textrm{from } c_3, \\
	& [-\tfrac{2}{3}, -\tfrac{2}{3}],		&& (-\tfrac{2}{3}, \infty), 	&& \textrm{from } c_4, \\
	& (-\infty, \underline{-14.5}),		&& [\underline{-14.5}, \underline{-14.5}] 	 	&& \textrm{from } c_5.
\end{align*}
We can select intervals $(-\infty, \tfrac{1}{2})$ and $(-1, \infty)$ to cover $\R$ and return this to the main call\footnote{The latter interval was provided by $c_2$, but we could have instead taken $(-\tfrac{2}{3}, \infty)$ from $c_4$ and still covered $\R$. \label{footnote-ex2origin}}.  

\begin{figure}[ht]
	\centering
        \includegraphics{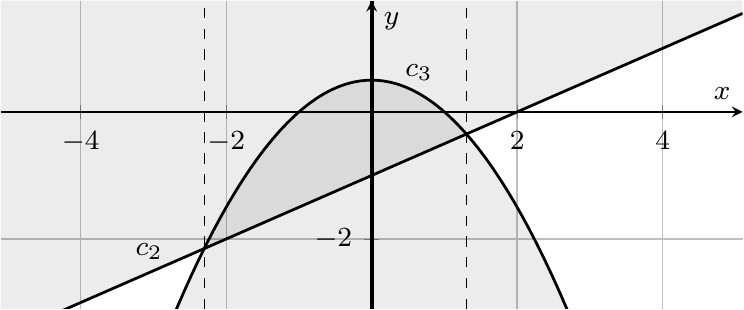}
        $\quad$
        \includegraphics{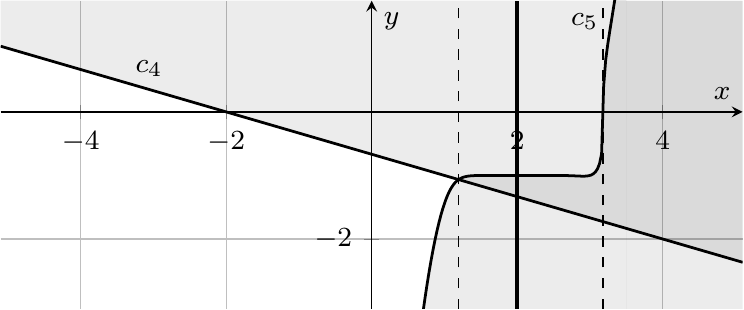}
	\caption{Excluded regions for $(x \mapsto 0)$ on the left and for $(x \mapsto 2)$ on the right.}\label{fig:example-2-detail}
\end{figure}

The main call analyses the covering and constructs the characterisation $\{ x^2 + x - 3 \}$ leading to the exclusion of the interval $(\underline{-2.30}, \underline{1.30})$. A graphical representation is shown in the left image of \cref{fig:example-2-detail}.
We then iterate through the loop again selecting a sample outside of this interval, say $s_1 = 2$, with which we enter another recursive call.

\paragraph{\texttt{get\_unsat\_cover($s = (x \mapsto 2)$)}}
This call is almost identical to the one above, only we are now forced to use $c_4$ to build a conflict with $c_3$ instead of $c_2$. The unsatisfiable intervals that initialise $\I$ are now as follows:
\begin{align*}
	& (-\infty, -58593751),	&& [-58593751, -58593751], && \textrm{from } c_1, \\
	& [0, 0], 			&& (0, \infty), 	&& \textrm{from } c_2, \\
	& (-\infty, -\tfrac{3}{2}),	&& [-\tfrac{3}{2}, -\tfrac{3}{2}], 	&& \textrm{from } c_3, \\
	& [\underline{-1.33}, \underline{-1.33}],	&& (\underline{-1.33}, \infty), && \textrm{from } c_4, \\
	& (-\infty, -1),	&& [-1, -1] 		&& \textrm{from } c_5.
\end{align*}
We obtain the (unique) minimal covering from these by selecting $(-\infty,-1)$ and $(\underline{-1.33},\infty)$ provided by $c_5$ and $c_4$ respectively. So we skip the loop and return this to the original parent call.

In the parent call the characterisation (after some simplification) contains two polynomials of degrees nine and eleven from the discriminant for $c_5$ and the resultant of the pair.  They each yield one real root: $\underline{1.19}$ from the resultant (corresponding to the crossing point of $c_4$ and $c_5$) and $\underline{3.18}$ from the discriminant corresponding to the point of inflection of $c_5$.   These are visualised in the right image of \cref{fig:example-2-detail}.  Note that the latter point is spurious in the sense that the point of inflection has no bearing on the satisfiability of our problem.  But while we could safely ignore it, we have no algorithmic way of doing so. Thus we exclude the interval $(\underline{1.19}, \underline{3.18})$ and proceed.

We now continue iterating the main loop with first the sample point $s_1 = \underline{3.18}$ and then $s_1 = 4$, which both yield the same conflict based on $c_5$ and $c_4$ and thus the exact same characterisation, excluding $[\underline{3.18},\underline{3.18}]$ and $(\underline{3.18},\infty)$ in turn. Finally we select $s_1 = -3$ and recurse one last time.

\paragraph{\texttt{get\_unsat\_cover($s = (x \mapsto -3)$)}}
Once again we compute the unsatisfiable intervals and obtain the following intervals in the initialisation of $\I$:
\begin{align*}
	& (-\infty, -1),	&& [-1, -1], 		&& \textrm{from } c_1, \\
	& [-\tfrac{5}{2}, -\tfrac{5}{2}],		&& (-\tfrac{5}{2}, \infty), 	&& \textrm{from } c_2, \\
	& (-\infty, -4),	&& [-4, -4], 		&& \textrm{from } c_3, \\
	& [\tfrac{1}{3}, \tfrac{1}{3}],		&& (\tfrac{1}{3}, \infty), 	&& \textrm{from } c_4, \\
	& (-\infty, -388),	&& [-388, -388] 	&& \textrm{from } c_5.
\end{align*}
We now use the unique covering $(-\infty,-1)$ and $(-\tfrac{5}{2},\infty)$ (from $c_1$ and $c_2$), yielding a characterisation consisting solely of the resultant of the polynomials defining $c_1$ and $c_2$ which has degree $11$. The excluded interval is $(-\infty,\underline{-2.06})$ which covers the whole region to the left of those excluded before.

At this point we have collected the following unsatisfiable intervals for the lowest dimension $x$ in our main function call.
\[
	(-\infty,\underline{-2.06}),
	(\underline{-2.30}, \underline{1.30}),
	(\underline{1.19}, \underline{3.18}),
	[\underline{3.18},\underline{3.18}],
	(\underline{3.18},\infty)
\]
We see that we have covered $\R$ and thus we return a final answer of UNSAT. Note that the highest degree of any polynomial we used in the above was eleven: the degree 33 resultant of $c_1$ and $c_5$ was never used nor even computed.

It is important to recognise the general pattern here that allows our algorithm to gain an advantage over a regular CAD. We have two more complicated constraints involved in creating the conflict, but they are separated in space, or at least the regions where they are needed to construct the conflict are separated. Increasing the degrees within $c_1$ and $c_5$ would affect the algorithm only insofar that we have to perform real root isolation on  larger polynomials while regular CAD has to deal with the resultant of these polynomials whose degree grows quadratically.

\subsubsection{Comparing to Incremental CAD}

The full projection contains one discriminant and 10 resultants (plus also some coefficients depending on which projection operator is used).  Since no satisfying sample will be found the incremental CAD will actually produce a full sign-invariant CAD for the polynomials in the problem, containing 273 cells.  

There is scope for some optimisations here, for example, because the constraints are all strict inequalities we know that they are satisfied if and only if they are satisfied on some full dimensional cell and so we could avoid lifting over any restricted dimension cells.  The CAD decomposes the real line into 27 cells and so we could optimise to lift only over the 13 intervals to consider 77 cells in the plane.  This avoids some work but we still need to compute and isolate real roots for the full projection set, and perform further lifting and real root isolation over half of the cells in $\R^1$.  This is significantly more real root isolation and even projection than computed by the new algorithm, in particular, the large resultant discussed above.

%Matthew: Figures in above paragraph are from CAD in Maple.  Gereon agreed with number of resultants and discriminants in projection.  But does your incremental CAD also end up computing 273 samples before concluding UNSAT? If not, how many?
%Gereon: I honestly don't know as SMT-RAT takes forever lifting a root of a degree-33 poly on another degree-11 poly. However given that the polynomials coincide I feel comfortable enough to just claim it. Improving our RANs for such cases is still on my todo-list, but will not happen this week.

%Full Projection for second example in SMT-RAT:
%\begin{align*}
%	coeffs(c_1) & degree(2), degree(9) \\
%	coeffs(c_2) & degree(1) \\
%	res(c_1, c_2) & degree(11) \\
%	coeffs(c_3) & degree(1), degree(1) \\
%	res(c_1, c_3) & degree(11) \\
%	res(c_2, c_3) & degree(2) \\
%	coeffs(c_4) & degree(1) \\
%	res(c_1, c_4) & degree(11) \\
%	res(c_2, c_4) & degree(1) \\
%	res(c_3, c_4) & degree(2) \\
%	disc(c_5) & degree(11) \\
%	coeffs(c_5) & degree(2), degree(9) \\
%	res(c_1, c_5) & degree(33) \\
%	res(c_2, c_5) & degree(11) \\
%	res(c_3, c_5) & degree(11) \\
%	res(c_4, c_5) & degree(11) \\
%\end{align*}

\subsubsection{Comparison to NLSAT}

SMT-RAT's implementation of NLSAT performs the following samples and conflicts for this example.
\begin{align*}
	x \mapsto 0 \quad & (c_2 \land c_3) \rightarrow (x \leq -2.3 \lor 1.3 \leq x) & (\underline{-2.3}, \underline{1.3}) \\
	x \mapsto -3 \quad & (c_1 \land c_2) \rightarrow (\underline{-2.06} \leq x) & (-\infty,\underline{-2.06}) \\
	x \mapsto 2 \quad & (c_4 \land c_5) \rightarrow (x \leq \underline{1.19} \lor x \geq \underline{3.18}) & (\underline{1.19}, \underline{3.184}) \\
	x \mapsto 5 \quad & (c_2 \land c_5) \rightarrow (x \leq \underline{3.199}) & (\underline{3.199},\infty) \\
	x \mapsto 52307/16384 \quad & (c_4 \land c_5) \rightarrow (x \leq \underline{3.184}) & (\underline{3.184},\infty) \\
	x \mapsto \underline{3.184} \quad & (c_4 \land c_5) \rightarrow (x \neq \underline{3.184}) & [\underline{3.184}, \underline{3.184}]
\end{align*}
Like the new algorithm it starts with the origin and gradually rules out the whole $x$-axis by generalising from a sample each time.  However, NLSAT required 6 iterations to do this, compared to 5 for the new algorithm.  The main reason for this is that it uses $c_2$ and $c_5$ (instead of $c_4$ and $c_5$) to explain the conflict at $x \mapsto 5$ and thus obtains a slightly smaller interval, leaving a gap between $\underline{3.184}$ and $\underline{3.199}$.  However, this poorer choice could have also been made by our new algorithm.

\subsubsection{Comparison to NuCAD}

\begin{figure}[ht]
	\begin{center}
	  \begin{subfigure}[b]{0.49\textwidth}
			\centering
          \includegraphics{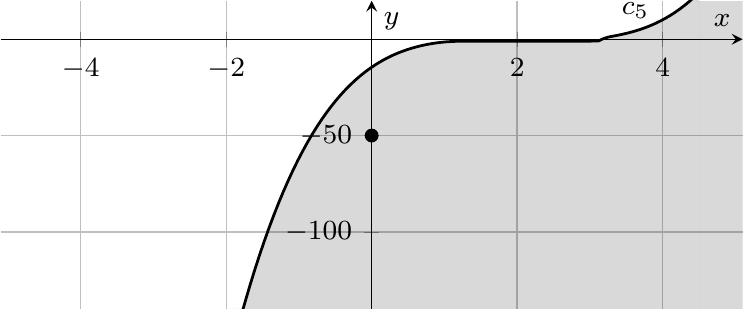}
			\caption{First NuCAD cell for $(0,-50)$.}\label{fig:example-compare-to-nucad:2}
			% Do not use \nucad here as this enforces \small
		\end{subfigure}
		\begin{subfigure}[b]{0.49\textwidth}
			\centering
                        \includegraphics{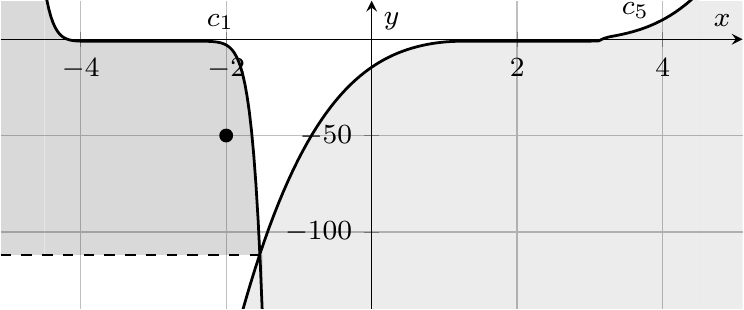}
			\caption{Second NuCAD cell for $(-2,-50)$.}\label{fig:example-compare-to-nucad:1}
			% Do not use \nucad here as this enforces \small
		\end{subfigure}
	\end{center}
		
	\caption{Graphs of the polynomials defining $c_1$ and $c_5$ in the more involved example (\ref{eq:second-example}).\label{fig:example-compare-to-nucad}}
\end{figure}

This time, because the example is UNSAT, NuCAD will have to complete and produce an entire decomposition of the plane.  The order in which cells are constructed will be driven by implementation as the only specification in the algorithm is to choose samples outside of cells with known truth value.  Since any reasonable implementation would start with the origin and prefer integer samples it is likely that NuCAD's computation path would follow similarly to our algorithm for this example.  But it should be noted that this is not guaranteed.  For example, if instead of the origin NuCAD were to start with the model $(0,-50)$ then here $c_5$ is violated and it may start by constructing the shaded cell in \cref{fig:example-compare-to-nucad:1}.

It must now pick a model outside of the shaded region.  Again, it would be a strange choice but the model $(-2, -50)$ would be acceptable and here $c_1$ is violated.  The necessary splitting would then require the calculation and real root isolation of the large resultant of the defining polynomials of $c_1$ and $c_5$.  There is one such real root, indicating a real intersection of the two polynomials as shown in \cref{fig:example-compare-to-nucad:2}, and this would necessarily be part of the boundary in the constructed cell.

We accept that the above model choices would be strange but they are a possibility for NuCAD while guaranteed to be avoided by the new algorithm.  Of course, it should be possible to shift the coordinate system of the example to make these model choices seem reasonable.

\subsection{Simple 3D Example}
\label{SSEC:3d1}

The 2D examples demonstrated how the new algorithm performs less work than a CAD and is more guided by conflict than NuCAD.  However, to demonstrate the benefits over NLSAT we need more dimensions.

\begin{figure}[t]
\includegraphics[width=0.48\textwidth]{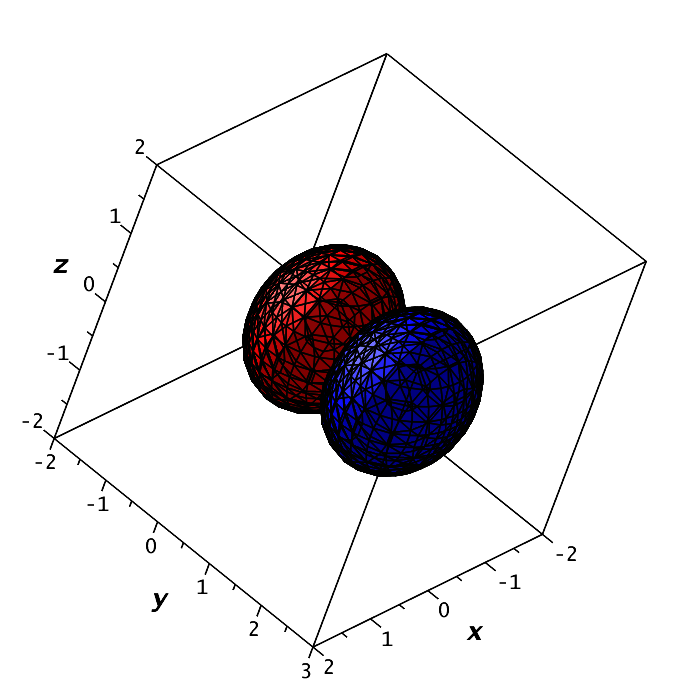}
\includegraphics[width=0.48\textwidth]{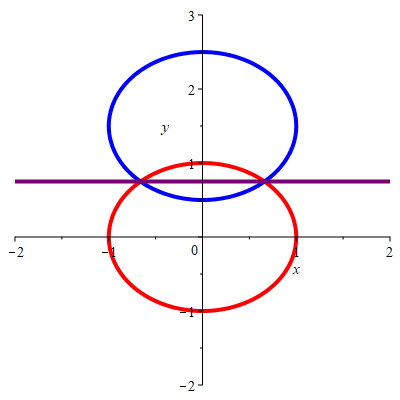}
\caption{On the left are the spheres defined by constraints (\ref{eq:3dS}) and on the right the cures of the $(x,y)$-plane computed as the first characterisation. \label{fig:simple3D}}
\end{figure}

Consider the simple problem in 3D of simultaneously satisfying the constraints:
\begin{equation}
c_1: x^2 + y^2 + z^2 < 1, \qquad c_2: x^2 + (y-\tfrac{3}{2})^2 + z^2 < 1. \label{eq:3dS}
\end{equation}
These require us to be inside two overlapping spheres as on the left of \cref{fig:simple3D}.  Let us consider how the new algorithm would find a witness.

\paragraph{\texttt{get\_unsat\_cover($s = ()$)}}
Our algorithm cannot draw any conclusions from the constraints as none are defined only in $x$, so it samples $x=0$ and enters the recursive call.

\paragraph{\texttt{get\_unsat\_cover($s = (x \mapsto 0)$)}}
Once again no conclusions can be drawn from the constraints yet so we sample $y = 0$ and recurse again.

\paragraph{\texttt{get\_unsat\_cover($s = (x \mapsto 0, y \mapsto 0)$)}}
This time the call of \texttt{get\_unsat\_intervals($s = (x \mapsto 0, y \mapsto 0)$)} does produce some conclusions.  It concludes that the first constraint is unsatisfiable outside of $x \in (-1,1)$ while the second constraint is not satisfiable anywhere over this sample.  I.e. the unsat interval $(-\infty,\infty)$ is part of the output.  This we may skip the main loop of \texttt{get\_unsat\_cover} and return to the previous call.

\paragraph{\texttt{get\_unsat\_cover($s = (x \mapsto 0)$) continued}}
For the characterisation we need calculate only the discriminant of the polynomial defining $c_2$, which after simplification is $x^2 + (y-\tfrac{3}{2})^2 -1$ (the blue circle in the right of \cref{fig:simple3D}).  We have no need to calculate the discriminant of the other defining polynomials, or their resultant (the other graphs on the right of \cref{fig:simple3D} which would be computed by a full CAD).  
%$y^2+x^2-1$, $4y^2 + 4x^2 - 12y + 5$ and $12y-9$ respectively. 

When forming the interval around $y=0$ we obtain the set $Z=\{-\infty, -1, \tfrac{1}{2}, \tfrac{5}{2}, \infty\}$ and so we generalise to $(-\infty, \tfrac{1}{2})$.  
\begin{itemize}
\item Sampling for $y$ anywhere outside $(\tfrac{1}{2}, 1)$ would lead to a full covering of the $z$-dimension after the initial querying of constraints in the recursive call obtained by analysing $c_1$.  The discriminant of $c_1$ would then be taken for the characterisation and the generalisation will rule out $(1, \infty)$.
\item Any sample from within $(\tfrac{1}{2}, 1)$ can be simply extended with $z=0$ to a full satisfying witness.
\end{itemize}

\subsubsection{Comparison to NLSAT}

NLSAT would proceed similarly in sampling $(x,y) = (0,0)$ and discovering the conflict.  However, it would then immediately build a cell around $(0,0)$ requiring the computation of the full projection of those polynomials in the conflict (i.e. not just the calculation of the discriminant above but then its discriminant also).  Only then would NLSAT move to a new partial sample.   In contrast, our new algorithm only computes projections with respect to the second variable once it has determined there is no possible $y$ to extend the $x$-value.  Since in this example there is such a $y$ for the first $x$-value those projections with respect to $y$ need never be computed.  Recall that iterated projection operations is the source of the doubly exponential growth in CAD, thus for a conflict which involved multiple constraints the savings are even more significant.

\subsection{More Involved 3D Example}

We finish with a larger 3D example to demonstrate some of the facets of the algorithm not yet observed in the smaller examples.  We define the three polynomials:
\begin{align}
f &:= -z^2 + y^2 + x^2 - 25,  \nonumber \\
g &:= (y-x-6)z^2 - 9y^2 + x^2-1, \nonumber \\
h &:= y^2-100, \label{eq:3d}
\end{align}
and seek to determine the satisfiability of
\[
f>0 \land g>0 \land h<0.
\]
We use variable ordering $x \prec y \prec z$, and note that unlike the previous example we have here not only a higher dimension, but also a non-trivial leading coefficient for $g$ and a constraint that is not in the main variable formed by $h$.  Finally please note that $z$ appears only as $z^2$ in the constraints and thus facts drawn for positive $z$ may be applied similarly for negative\footnote{We are not suggesting our algorithm makes this simplification, we just seek to reduce the presentation of details here.}.

\begin{figure}[p]
\includegraphics[width=0.495\textwidth]{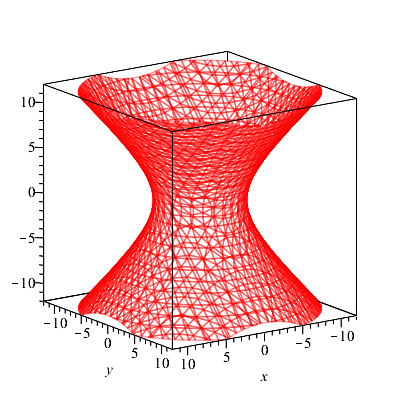}
\includegraphics[width=0.495\textwidth]{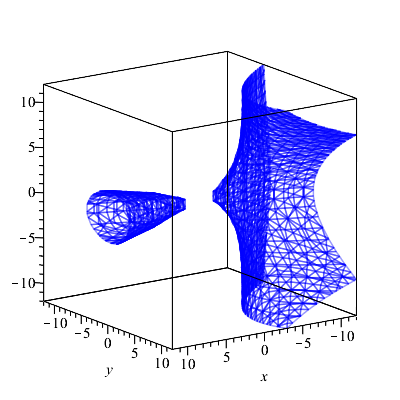}
\includegraphics[width=0.495\textwidth]{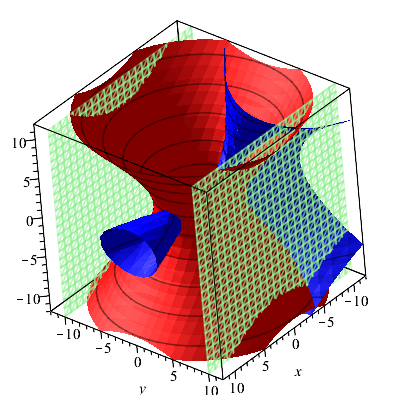}
\includegraphics[width=0.495\textwidth]{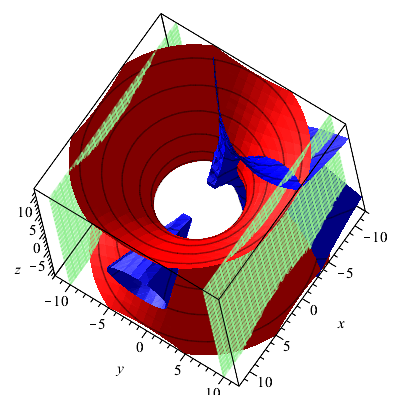}
\caption{Surfaces defined by the polynomials in (\ref{eq:3d}). \label{fig:3d}}
\end{figure}

The surfaces defined by the polynomials are visualised in \cref{fig:3d} where the red surface is for $f$, the blue for $g$ and the green for $h$.  We see that $f$ is a hyperboloid, while $g$ would have been a paraboloid if it were not for the leading coefficient in $z$.  The final constraint simply bounds $y$ to $(-10,10)$.  We see that there are many non-trivial intersections (and self intersections) between the surfaces.  A full sign-invariant CAD for the three polynomials may be produced with the Regular Chains Library for Maple \cite{CM14a} in about 20 seconds, and contains 3509 cells.

\begin{figure}[ht]
  \centering
  \includegraphics{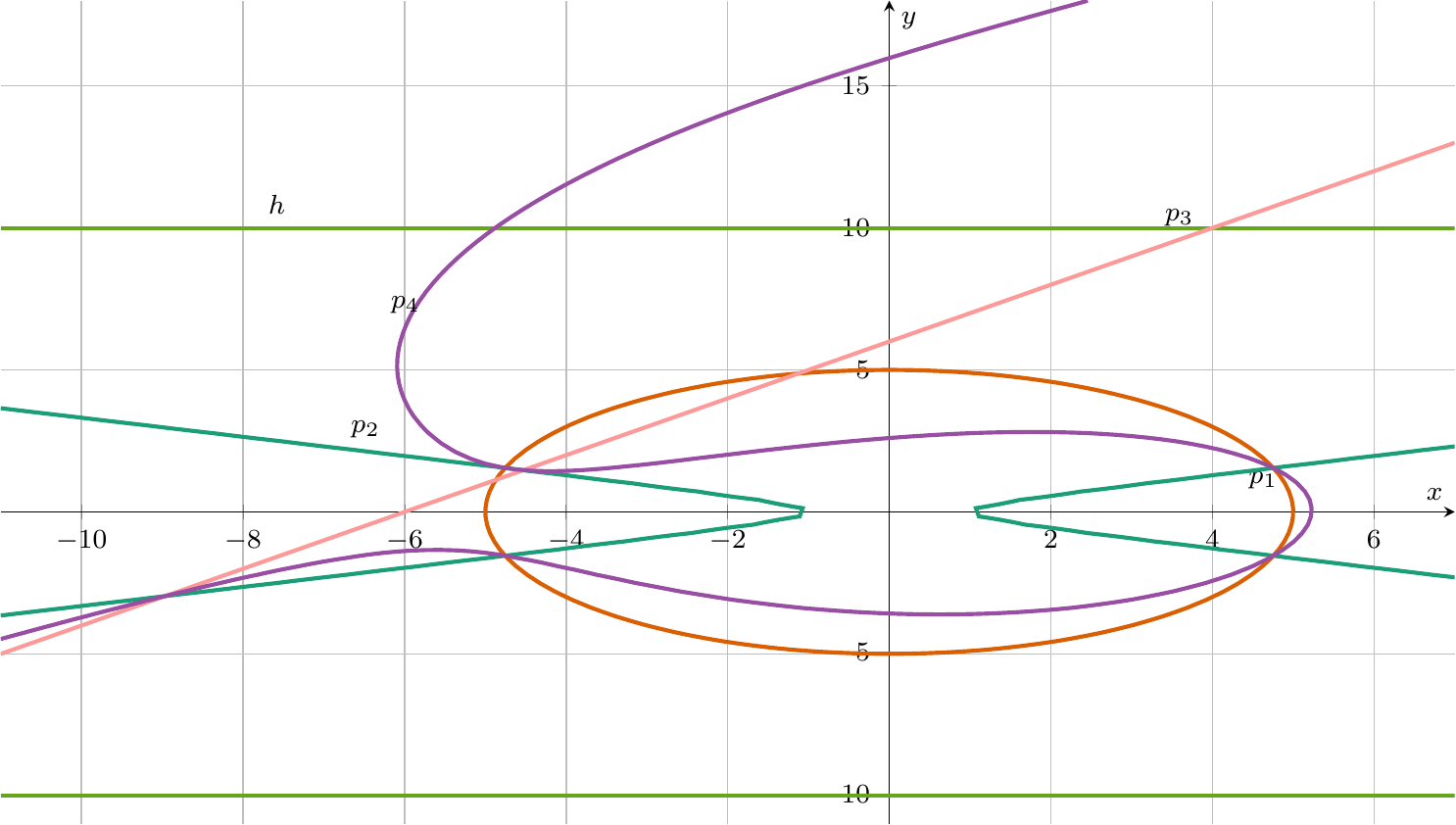}
	\caption{Projection of the surfaces in \cref{fig:3d}. \label{fig:3dProj2}}
\end{figure}

For use in the discussion later, let us examine and put a label on all the CAD projection polynomials in $(x,y)$ (i.e. those obtained after $z$ is projected out)\footnote{We do this first just to ease presentation: the algorithm would only compute a projection polynomial when it needed it.}:
\begin{align*}
p_1 &:= y^2+x^2-25 \\
p_2 &:= -9y^2+x^2-1 \\
p_3 &:= -y+x+6 \\
p_4 &:= -y^3 + y^2x + 15y^2 - yx^2 + 25y + x^3 + 5x^2 - 25x -149
\end{align*}
Here $p_1$ is the discriminant of $f$ with respect to $z$ (at least up to a constant factor); $p_2$ and $p_3$ are factors of the discriminant of $g$ and also the trailing and leading coefficients of $g$ respectively.  The resultant of $f$ and $g$ with respect to $z$ evaluated to $p_4^2$.    Of course, the projection also includes $h$ itself and is shown in \cref{fig:3dProj2}.

Also for use in the following discussion, we label the six real roots in $y$ that these four polynomials have when evaluated at $x=0$:
\begin{align*}
b_1 &= -5 \mbox{ defined by } p_1 \\
b_2 &= \underline{-3.58} \mbox{ defined by } p_4 \\
b_3 &= 2.60 \mbox{ defined by } p_4 \\
b_4 &= 5 \mbox{ defined by } p_1 \\
b_5 &= 6  \mbox{ defined by } p_3 \\
b_6 &= \underline{15.98} \mbox{ defined by } p_4 
\end{align*}
Note that $p_2$ has no real roots when $x=0$.%, and that the three roots stated as decimals are given only to two decimal places and actually stored of algebraic numbers (the unique roots of $p_4$ within a bound).  

%\begin{figure}[ht]
%\centering
%\includegraphics[width=0.5\textwidth]{3DExProj.png}
%\caption{Projection of the surfaces in \cref{fig:3d}). \label{fig:3dProj}}
%\end{figure}

Now let us describe how the new algorithm would determine satisfiability.

\paragraph{\texttt{get\_unsat\_cover($s = ()$)}}
Our algorithm cannot draw any conclusions from the constraints originally as none are defined only in $x$, so it samples $s_1=0$ and enters the recursive call.

\paragraph{\texttt{get\_unsat\_cover($s = (x \mapsto 0)$)}}
This time, the call of \texttt{get\_unsat\_intervals($s = (x \mapsto 0)$)} does obtain some information from the third constraint.  It produces two unsat intervals: $(-\infty, -10]$ and $[10, \infty)$ both of which are characterised by $h$.  However, since this does not cover the whole line we enter the main loop and sample outside, starting with $s_2=0$.

\paragraph{\texttt{get\_unsat\_cover($s = (x \mapsto 0, y \mapsto 0)$)}}
This time the call of \texttt{get\_unsat\_intervals($s = (x \mapsto 0, y \mapsto 0)$)} produces two intervals, both of which are $(-\infty, \infty)$ defined by $f$ and $g$ respectively.  Thus the main loop is skipped and we return to the prior call.

\paragraph{\texttt{get\_unsat\_cover($s = (x \mapsto 0)$) continued}}
For the characterisation we need pick only one of the two intervals in $\I$ since both cover the whole line (in the call to \texttt{construct\_cover} in Algorithm \ref{alg:construct-characterization}).  Let us suppose we pick the one formed by $g$.  Then the characterisation returns $\{p_2, p_3\}$.  When forming the interval around $y=0$ we obtain the set $Z=\{-\infty, 6, \infty\}$ and so we generalise to $(-\infty, 6)$ with the bound defined by $p_3$.  We still do not have a full covering so we loop again.  The uncovered space is now $[6, 10)$ so we must sample from here.  Suppose we sample $s_2=7$ and recurse.

\paragraph{\texttt{get\_unsat\_cover($s = (x \mapsto 0, y \mapsto 7)$)}}
This time the initial constraints do provide a cover but only when considered together: $f$ requires $z \in (\underline{-4.90}, \underline{4.90})$ while $g$ requires $z \notin [\underline{-21.02}, \underline{21.02}]$.%\footnote{Again, these floats represent actual algebraic numbers: $2\sqrt{6}$ in the former and $\sqrt{442}$ in the latter.}.

\paragraph{\texttt{get\_unsat\_cover($s = (x \mapsto 0)$) continued}}
This time the characterisation contains all of $\{p_1, p_2, p_3, p_4\}$ and so $Z$ contains all of the $b_i$.  Hence we can generalise around $x=7$ to the interval $(6, \underline{15.98})$.  We now have a cover for all $y$ values over $x=0$ except $y=6$.

\paragraph{\texttt{get\_unsat\_cover($s = (x \mapsto 0, y \mapsto 6)$)}}
This time a complete cover is provided from the initial constraint defined by $g$ so no need for the main loop.

\paragraph{\texttt{get\_unsat\_cover($s = (x \mapsto 0)$) continued}}
Thus the characterisation is similar to that obtained from the sample $y=0$.  We actually have to do a little extra work here:  because at $(x,y)=(0,6)$ the leading coefficient of $g$ is zero we include also the trailing coefficient, but this was actually already a factor of the discriminant so we do have the same characterisation as for $y=0$.  Because this characterisation produced a $Z$ containing $y=6$ we cannot generalise beyond this point.  Nevertheless, we have a full cover over $x=0$.

\paragraph{\texttt{get\_unsat\_cover($s = ()$) continued}}
The cover is formed from five intervals: two from the initial constraints and three from the recursions as summarised below.

\begin{center}
\begin{tabular}{ccccccc}
 & $l$ 		& $u$ 		& $L$ 		& $U$ 		& $P_i$		& $P_{\bot}$ 
 \\ \hline
 & $-\infty$ 	& $-10$		& $\emptyset$ 	& $\{h\}$ 	& $\{h\}$ 	& $\emptyset$ 
 \\
$I_1$ & 
$-\infty$	& $6$		& $\emptyset$	& $\{p_3\}$	& $\{p_2,p_3\}$ & $\emptyset$ 
\\
$I_2$ & 
$6$			& $6$		& $\{p_3\}$	& $\{p_3\}$	& $\{p_2,p_3\}$ & $\emptyset$ 
\\
$I_3$ &
$6$			& $\underline{15.98}$	& $\{p_3\}$	& $\{p_4\}$	& $\{p_1,p_2,p_3,p_4\}$ & $\emptyset$ 
\\
$I_4$ &
$10$ 		& $\infty$	& $\{h\}$ 	& $\emptyset$ 	& $\{h\}$ 	& $\emptyset$ 
\end{tabular}
\end{center}

The first of these intervals is redundant: the interval $(-\infty,-10)$ from $h$ is entirely inside $(-\infty, 6)$ from the sample $y=0$.  Hence it will be removed by the call to \texttt{construct\_cover}, where we also order the remaining four intervals as in the table above.   

We now form the characterisation at $x=0$: it contains the discriminants of $p_1, p_2$ and $p_4$; those of $p_3$ and $h$ were constant and so removed. Similarly, all the leading coefficients were constant and so not required.
The only within interval resultants required come from $I_3$.  We take the resultant of $p_3$ with both $p_1$ and $p_4$ as both the latter produce smaller real roots. The only between interval resultant required is that between $p_4$ and $h$ which protects the overlap of the final two intervals.  After the standard simplifications (including factorisation) we have the following characterisation.
\begin{align*}
R &:= \{
x-5, x+5, x-1, x+1, 4x^6+20x^5+475x^4+6760x^3+7670x^2-198300x-655237, \\
&\qquad 2x^2+12x+11, 8x^2+108x+325, x^3-5x^2+75x+601, x^3+15x^2+75x+2101
\}\ .
\end{align*}
This produces the following set of real roots:
\[
Z = \{ 
\underline{-17.55}, \underline{-8.97}, \underline{-6.09}, -5, \underline{-4.88}, \underline{-4.87}, \underline{-4.53}, \underline{-1.13}, -1, 1, 5, \underline{5.23}
\}\ .
\]
%where once again the floating point numbers are just decimal approximations to real algebraic numbers stored in the algorithm.  
Hence we can generalise around $x=0$ to the interval $(-1,1$).

\begin{figure}[ht]
	\centering
	\begin{subfigure}[b]{0.49\textwidth}
		\centering
                \includegraphics{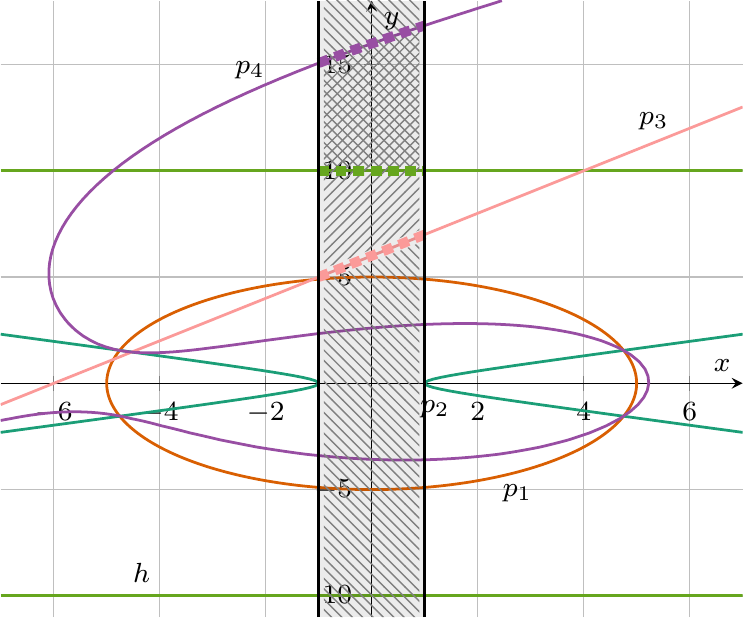}
		\caption{The cover over $x \in (-1,1)$ generalised from $x=0$. \label{fig:3dCover2}}
	\end{subfigure}
	\begin{subfigure}[b]{0.49\textwidth}
	  \centering
          \includegraphics{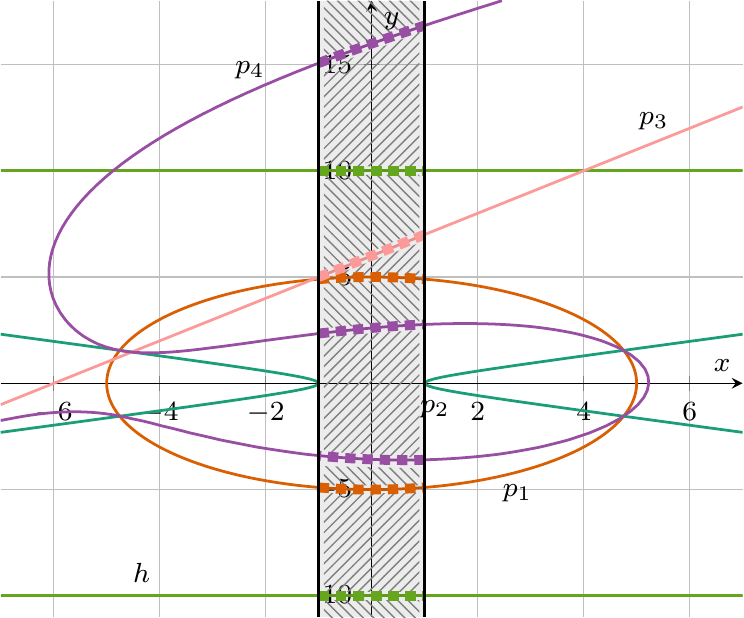}
		\caption{The cylinder over $x \in (-1,1)$ as produced by CAD. \label{fig:3dCover-cad-cylinder}}
	\end{subfigure}
	\caption{Comparison of the cover over $x \in (-1,1)$ with a cylinder from regular CAD.}
\end{figure}

\cref{fig:3dCover2} visualises this generalisation of the cover.  We see that $I_1$, which was originally the $y$ axis below $6$, is generalised to the cylinder bounded at the top by $p_3$; while $I_2$, which was originally just the point $(x,y)=(0,6)$, is generalised to the line segment of $y-x-6$ that is above $x \in (-1,1)$.   Recall that $I_2$ was a cell where the leading coefficient of $g$ vanished and observe that this remains the case throughout the generalisation.  $I_3$ and $I_4$ were overlapping segments of the $y$-axis and now form overlapping cylinders.  These different cells are indicated by the different direction of shading in the figure.

Let us pause to compare this cylindrical algebraic covering with an actual CAD.  A CAD of the projection $\{p_1, p_2, p_3, p_4, h\}$ would also form a cylinder over $x \in (-1,1)$.  However, this cylinder would then be split into 17 cells according to the 8 line segments in the image.  We would have to work over 17 samples to determine UNSAT over the cylinder, while we obtained the same conclusion using only 4 samples in the new algorithm.

The algorithm continues in the call to \texttt{get\_unsat\_cover($s = ()$)} with a sample outside of $(-1,1)$.  A natural choice here is $s_1=2$, for which an UNSAT cover would be found consisting of the following six intervals (after removal of redundant ones):

\begin{center}
\begin{tabular}{ccccccc}
 & $l$ 		& $u$ 		& $L$ 		& $U$ 		& $P_i$		& $P_{\bot}$ 
 \\ \hline
$I_1$ & 
$-\infty$	& $\underline{-0.58}$	& $\emptyset$	& $\{p_2\}$	& $\{p_2,p_3\}$ & $\emptyset$ 
\\
$I_2$ & 
$\underline{-4.58}$		& $\underline{4.58}$	& $\{p_1\}$	& $\{p_1\}$	& $\{p_1\}$ & $\emptyset$ 
\\
$I_3$ &
$\underline{0.58}$		& $8$		& $\{p_2\}$	& $\{p_3\}$	& $\{p_2,p_3\}$ & $\emptyset$ 
\\
$I_4$ &
$8$ 		& $8$		& $\{p_3\}$ & $\{p_3\}$	& $\{p_2,p_3\}$ 	& $\emptyset$ 
\\
$I_5$ &
$8$ 		& $\underline{17.64}$	& $\{p_3\}$ & $\{p_4\}$	& $\{p_1,p_2,p_3,p_4\}$ 	& $\emptyset$ 
\\
$I_6$ &
$10$ 		& $\infty$	& $\{h\}$ & $\emptyset$	& $\{h\}$ 	& $\emptyset$ 
\end{tabular}
\end{center}

The characterisation then allows a generalisation from $x=2$ to $x \in (1, 4.75)$, promoting a third sample in the main call of $x=5$.  In the recursion a sample of $y=0$ is unsat for the first constraint but cannot be generalised.  A second sample of $y=1$ leads to the initial constraints providing a cover for all but $z \in (-1,1)$.  Hence the main loop is entered and we sample $z=0$ to find a satisfying witness:  $(x,y,z) = (5,1,0)$ which evaluates $f,g,h$ to $1, 15$ and $-99$ respectively.

%\subsubsection{Comparison}
%Incremental CAD (with the same variable ordering) also finds the same assignment $(5,1,0)$ but needs less polynomials to do so.  It only computes $disc(f), disc(g), lcoeff(g), res(f,g), res(h,p_3)$.  If we reverse the variable order ($z,y,x$) it gets even ``worse'': there is not a single additional polynomial is computed and it doesn't even look at $g$ to find $(6,0,0)$.  However, this is due to luck rather than guidance.
%NLSAT is a bit difficult right now because we are currently investigating dynamic variable ordering. With the variable ordering $y,x,z$ NLSAT excludes only two cells.
% MATTHEW agrees that there is no need to include this.

We note that the covering produces far less cells than a  traditional CAD.  In particular, not every projection factor is used in every covering.  This was demonstrated visually by \cref{fig:3dCover2} where the purple resultant was a cell boundary for positive $y$ but not for negative $y$.

\section{Experiments on the SMT-LIB}
\label{SEC:Experiments}

\subsection{Aim of these Experiments}

We note at the outset that the purpose of these experiments is to establish which of the different algorithmic ideas discussed throughout the paper performs best on a substantial dataset of real problems.  The aim is not to prove that a particular solver is the current state of the art.  Given our aim it made sense to use implementations of the algorithms in a single system so that they can all draw on the same underlying data-structures and sub-algorithms.  This way we can more fairly compare the effects of the high level algorithm with less risk of effects from implementation issues.  Thus all the competing solvers in this section are implemented in the established SMT solver SMT-RAT~\cite{smtrat}.

\newcommand{\SolverCADNaive}{\texttt{CAD-naive}\xspace}
\newcommand{\SolverCADFull}{\texttt{CAD-full}\xspace}
\newcommand{\SolverCDCAD}{\texttt{CDCAC}\xspace}
\newcommand{\SolverMCSAT}{\texttt{NLSAT}\xspace}

\subsection{Dataset}

The worked examples show that the new algorithm does have advantages over existing techniques.  We will now check whether this translates into a method that is also competitive in practice.   This section describes experiments on the benchmarks for quantifier-free non-linear real arithmetic (\texttt{QF\_NRA}) from the SMT-LIB~\cite{Barrett2016} initiative (as of March 2020). This dataset contains $11489$ problem instances from $11$ different families.  Note that these examples are not presented as sets of constraints but can come with more involved Boolean structure, and so the new algorithm is applied as the theory solver for the larger SMT system.  Throughout we apply a time limit of $60$ seconds and a memory limit of 4GB for tackling a problem instance.

\subsection{New Solver \SolverCDCAD}

We produced an implementation in the course of a Master's thesis \cite{Franzen2020} within SMT-RAT.  The solver \SolverCDCAD consists of the regular SAT solver from SMT-RAT, combined with this implementation of the presented method as the only theory solver.  
We consider this implementation to be a proof of concept, as it does not implement any form of incrementality and there is scope to optimise several subroutines.  In the interests of reproducibility it is available from the following URL:
\begin{center}
\url{https://github.com/smtrat/smtrat/tree/JLAMP-CDCAC}
\end{center}

With regards to the completeness issue discussed in Section \ref{SSEC:Completeness}: we observed nullification upon substitution of the sample (Algorithm \ref{alg:interval-from-characterization} Line \ref{realroots}) in a small, but significant number of examples: $82$ from all $11489$ instances (0.7\%) involved a nullification within the time limit.  
This means that for these instances the theory does not guarantee UNSAT results of solver \SolverCDCAD as correct.  However, we note that in all cases the results provided were still correct (i.e. they matched the declared answer stored in the SMT-LIB), which is consistent with our experience of similar CAD completeness issues in~\cite{Kremer2019}.

\subsection{Competing Solvers}
\label{SSEC:OtherSolvers}

We compare the new algorithm to a theory solver based on traditional CAD as described in~\cite{Kremer2019}.  We show this in two configurations: \SolverCADNaive uses no incrementality (by projection factor $-$ see Section \ref{sec:related}) while \SolverCADFull employs what was called \emph{full incrementality} in~\cite{Kremer2019}.  Solver \SolverCDCAD uses no incrementality, as already discussed in Section \ref{SUBSEC:TheorySolver}, but the difference between the incremental and non-incremental versions of the full CAD gives a glimpse at what further improvements to \SolverCDCAD might be possible.  Both of these use Lazard's projection operator.

We also compare against the NLSAT method~\cite{JovanovicdeMoura2012a}. For better comparability of the underlying algorithms, we used our own implementation of the NLSAT approach which uses the same underlying data structures as the aforementioned solvers.  NLSAT uses only the CAD-based explanation backend and the NLSAT variable ordering.  This uses the same projection operator as \SolverCDCAD and so has similar completeness limitations.  We call this solver \SolverMCSAT and refer to~\cite{Nalbach2019} for a more detailed discussion of the implementation and the relation to the original NLSAT method from~\cite{JovanovicdeMoura2012a}.  

\subsection{Absence of NuCAD}

Throughout the paper we also highlighted NuCAD as a relevant competing algorithm.  NuCAD is not currently implemented in SMT-RAT and so could not be included directly in the experimental evaluation.  There is only a single implementation of NuCAD in existence:  in the \texttt{Tarski} system~\cite{Brown2017}.  However, \texttt{Tarski} only implements \emph{open NuCAD} (full dimensional cells only) which reduces the comparability, and there are also some technical issues at the time of writing with \texttt{Tarski}'s parsing of certain SMT-LIB benchmarks and so we are not able to report on the use of \texttt{Tarski} here.  We hope, however, that the worked examples and discussions above made sufficiently clear the differences and advantages of the new algorithm in comparison to NuCAD.

\subsection{Determinism and Scrambling}

We stress that the algorithms compared are all deterministic.  I.e. if we ran the benchmark experiments again we would get the exact same results. Furthermore, though we did not try experimentally, we do not expect that a random scrambling of the benchmarks would cause any major change in the general observations. The only potential effect we can foresee would be to make different \emph{heuristical} choices. For example, a renaming of the variables could cause the solver to pick a different choice in situations where the decision is based on the lexicographic order of the variables.
However, if the algorithms that we compare share the same heuristics then scrambling might bring advantages or disadvantages on single problem instances,  but it should not cause any fundamental changes visible on larger benchmark sets.

\subsection{Experimental Results}

\begin{figure}[ht]

  \begin{tabularx}{\linewidth}{lX
S[table-format=4.0]@{~}S[table-format=2.2]@{\,}s
S[table-format=4.0]@{~}S[table-format=2.2]@{\,}s
S[table-format=5.0]@{~}S[table-format=2.1]@{\,}s}
	\toprule
	\textbf{Solver} &
	& \multicolumn{3}{c}{\textbf{SAT}}
	& \multicolumn{3}{c}{\textbf{UNSAT}}
	& \multicolumn{3}{c}{\textbf{overall}}
	\\
	\midrule
	\SolverCADNaive{} &
	& 4277 & 0.24 & \second{}
	& 3407 & 0.95 & \second{}
	& 7684 & 66.9 & \percent{}
	\\
	\SolverCADFull{} &
	& 4309 & 0.20 & \second{}
	& 4207 & 1.50 & \second{}
	& 8516 & 74.1 & \percent{}
	\\
	\SolverCDCAD{} &
	& 4365 & 0.50 & \second{}
	& 4373 & 1.20 & \second{}
	& 8738 & 76.1 & \percent{}
	\\
	\SolverMCSAT{} &
	& 4521 & 0.40 & \second{}
	& 4409 & 1.32 & \second{}
	& 8930 & 77.7 & \percent{}
	\\
	\bottomrule
  \end{tabularx}
  
	\caption{Experimental results for different CAD-based solvers.}\label{tbl:experiments}
\end{figure}

\begin{figure}[hb]
	\centering
          \includegraphics{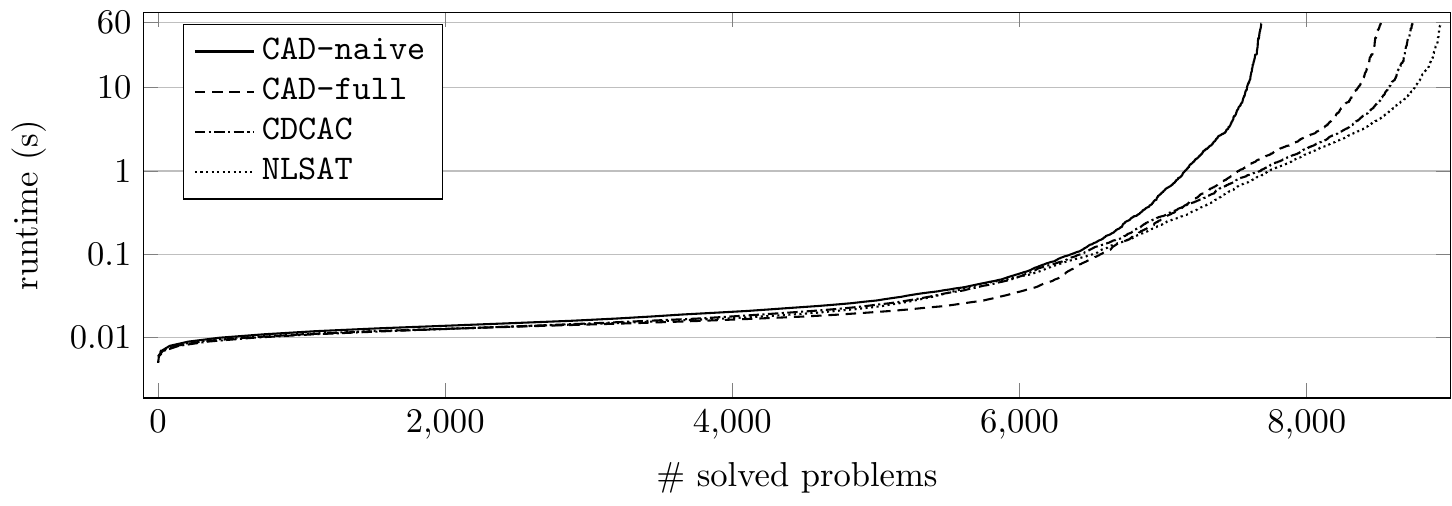}
	\caption{Graphical depiction of the experimental results for different CAD-based solvers.}\label{fig:experiments}
\end{figure}

The overall performance of the solvers is summarised in \cref{tbl:experiments}.  For each solver we show the number of solved problem instances (within the 60 second time limit), both separated into satisfiable and unsatisfiable instances and in total. Additionally, average runtimes for the solved instances are given and the final column shows the overall percentage of the dataset that was solved.  
The table shows that our new algorithm has led to a competitive solver.  It significantly outperforms a solver based on regular CAD computation, even when it has been made incremental.   \cref{fig:experiments} shows how the solvers tackled instances by their runtime, demonstrating that on trivial examples the regular CAD actually outperforms the others, but for more complex examples the savings of the adapted solvers outweigh their overheads.

We note that although we use the same SAT solver for \SolverCADNaive, \SolverCADFull and \SolverCDCAD, the overall SMT solver is not guaranteed to behave the same, i.e. make the same sequence of theory calls.  This is because the theory solvers may construct different infeasible subsets which guide the SAT solver in different ways. Hence the difference seen in \cref{tbl:experiments} may not be purely due to improved performance within the theory solver.  Our prior experience with this issue from~\cite{Hentze2017,Kremer2019} indicates that the different subsets usually have only a negligible influence on the overall solver performance.
%It should be evaluated in the future whether this holds true for the novel method as well.

\subsection{Comparison with \nlsat in Experimental Results}

We acknowledge that our new algorithm is outperformed by the \nlsat-based solver.  We note that this out-performance is greater on SAT instances than UNSAT, suggesting that the new approach may be particularly useful for proving UNSAT.  
We are optimistic that \SolverMCSAT could be beaten following improvements to our implementation.  First, implementing incrementality increased the number of problems the CAD solver could tackle by 7\% of the dataset, so gains to \SolverCDCAD from incrementality may be particularly fruitful.  Second, we have yet to perform optimisation on things like the variable ordering choice for \SolverCDCAD, which is known to have a great effect on CAD-based technology (see e.g. \cite{EF19}) and has been already performed for this \nlsat-based solver to give significant improvements as detailed in ~\cite{Nalbach2019}.  

\begin{figure}[ht]
	\begin{subfigure}[b]{0.6\textwidth}
          \includegraphics{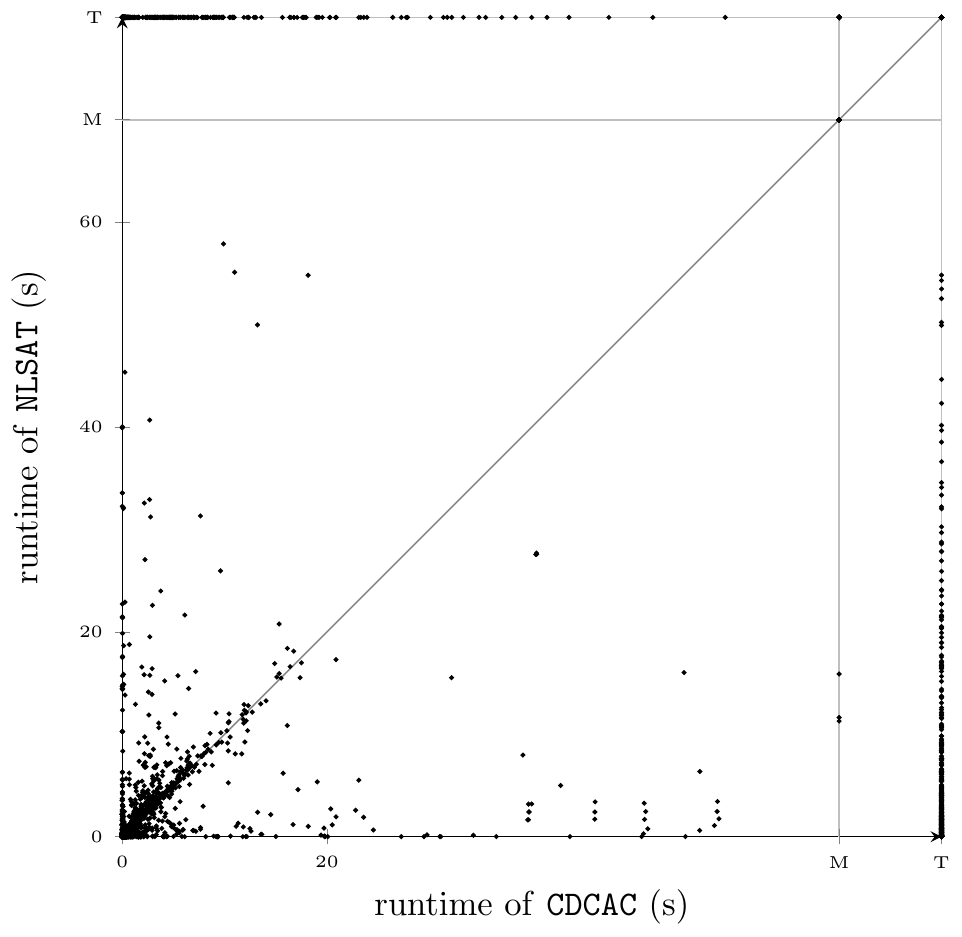}
	  \caption{Scatter plot of \SolverCDCAD and \SolverMCSAT.  Here $M$ and $T$ signify respectively that the 4GB memory limit or 60s time limit was met for an instance.}\label{fig:scatter}
	\end{subfigure}
	\begin{subfigure}[b]{0.39\textwidth}

	  \begin{tabularx}{\textwidth}{Xrr}
	\multicolumn{3}{c}{\# Problems for which \SolverCDCAD is} \\
	& SAT & UNSAT \\
at least $10\times$ slower & 422 & 322 \\
at least $2\times$ slower & 822 & 628 \\
at least $1.1\times$ slower & 1907 & 1374 \\
similar & 922 & 1524 \\
at least $1.1\times$ faster & 1778 & 1743 \\
at least $2\times$ faster & 640 & 566 \\
at least $10\times$ faster & 197 & 285 \\

          \end{tabularx}
          
		\caption{Comparison with \SolverMCSAT for runtimes $\geq 10ms$.}\label{fig:compare-runtimes-10ms}

		\vspace*{1cm}

\begin{tabularx}{\textwidth}{Xrr}
	\multicolumn{3}{c}{\# Problems for which \SolverCDCAD is} \\
	& SAT & UNSAT \\
at least $10\times$ slower & 422 & 322 \\
at least $2\times$ slower & 599 & 546 \\
at least $1.1\times$ slower & 629 & 704 \\
similar & 28 & 823 \\
at least $1.1\times$ faster & 395 & 808 \\
at least $2\times$ faster & 356 & 430 \\
at least $10\times$ faster & 197 & 284 \\

\end{tabularx}

		\caption{Comparison with \SolverMCSAT for runtimes $\geq 100ms$.}\label{fig:compare-runtimes-100ms}
	\end{subfigure}
	\caption{Direct comparison of \SolverCDCAD and \SolverMCSAT.}\label{fig:direct-comparison}
\end{figure}

A more detailed comparison of \SolverCDCAD and \SolverMCSAT is presented in \cref{fig:direct-comparison}.  
In \cref{fig:scatter} we show a direct comparison of the two solvers, where every dot represents one problem instance.
The cluster of points in the lower left corner represent the many \emph{easy} problem instances that are solved by both solvers and thus is not so important for the comparison.  More interesting is that the solvers perform very differently on a substantial amount of harder problem instances.  Note that almost no problem instance is \emph{solved by both} solvers after more than $20$ seconds, but that both solvers can solve such problem instances \emph{that the other cannot}.  There are 555 problems on which \SolverCDCAD times out but NLSAT completes, and 358 problems for which NLSAT times out and \SolverCDCAD completes.  

\cref{fig:compare-runtimes-10ms,fig:compare-runtimes-100ms} separates the number of problem instances according to how much slower or faster \SolverCDCAD is when compared to \SolverMCSAT. The data in \cref{fig:compare-runtimes-10ms} are for problems where at least one solver took more than $10ms$, and the data in \cref{fig:compare-runtimes-100ms} where at least one took above $100ms$ (so these exclude instances that the solvers agree are trivial).  Note that the rows are cumulative (i.e. instances which are $2\times$ faster are also $1.1\times$ faster).  \cref{fig:compare-runtimes-10ms} states that \SolverMCSAT was actually slower than \SolverCDCAD more often than it was faster!  This does not contradict the data in \cref{tbl:experiments} however, as \SolverMCSAT still completed in many of the cases it was slower while \SolverCDCAD would time out more often when slower.  We looked for whether the stronger performance of a solver correlated to different origins of the problem instances but did not find strong evidence of this.

Altogether, the data from \cref{fig:direct-comparison} indicates that \SolverCDCAD and \SolverMCSAT have substantially different strengths and weaknesses.  There should hence be benefit in combining them, for example in a portfolio solver \cite{XHHL08}.

\section{Conclusions}
\label{SEC:Conc}

We have presented a new algorithm for solving SMT problems in non-linear real arithmetic based on the construction of cylindrical algebraic coverings to learn from conflicts.  We demonstrated the approach with detailed worked examples and the performance of our implementation on a large dataset.

\subsection{Comparison with Traditional CAD}

There are clear advantages over a more traditional CAD approach (even one that is made incremental for the SMT context).  The new algorithm requires fewer projection polynomials, fewer cells, and fewer sample points to determine unsatisfiability in a given cylinder.  It can even sometimes avoid the calculation of projection polynomials entirely, when they represent combinations of polynomials not required to determine unsatisfiability.  These advantages manifested in substantially more examples solved in the experiments.

\subsection{Comparison with \nucad}

The worked examples made clear how the new algorithm more effectively learns from conflicts than the \nucad approach \cite{Brown2015} applied to our setting.  Although \nucad also saves in projection and cell construction when compared to traditional CAD, its savings are not maximised as the search is less directed.  The conflicts of separate cells in \nucad are never combined and so the search within \nucad is not guided to the same extent.  However, \nucad is able to work in a larger QE context, while this new algorithm is applicable only in a restricted setting for checking the consistency of polynomial constraint sets, as outlined in Section \ref{SSEC:Setting}.

\subsection{Comparison with \nlsat}

Our ideas are closest to those of the \nlsat method~\cite{JovanovicdeMoura2012a} which likewise builds partial samples and when these cannot be extended explains the conflict locally and generalises from the point.  For several of the worked examples the new algorithm and \nlsat performed similarly, with any differences due to luck or heuristic choices not fundamental to the theory.  However, the methods are very different in their presentation and computational flow.  There are a number of potential advantages of the new approach compared to \nlsat.  

First, on a practical note, our method is SMT compliant and thus allows for a (relatively) easy integration with any existing SMT solver. \nlsat on the other hand is usually separate from the regular SMT solving engine which makes it hard to combine with other approaches, e.g. for theory combination.  Second, our approach can avoid some projections performed by \nlsat in cases where a change in the sample in one dimension will find a satisfying witness (see the example in Section \ref{SSEC:3d1}).  

Most fundamentally, our approach keeps the theory reasoning away from the SAT solving engine while \nlsat burdens it with many lemmas for the (intermediate) theory conflicts. For a problem instance that is fully conjunctive our method may determine the satisfiability alone while NLSAT will still require a SAT solver.  
It is possible that NLSAT may cause the Boolean reasoning to slow down or use excessive amounts of memory by forcing it to deal with a large number of clauses.  Further, once we move on to a different area of the solution space most of the previously generated lemmas are usually irrelevant, but their literals may still leak into the theory reasoning in \nlsat.  An interesting claim about \nlsat is that explanations for a certain theory assignment can sometimes be reused in another context \cite{deMoura_MCSAT}. While this may be the case, our own experiments suggest that this impact is rather limited. Our new algorithm can not reuse unsatisfiable intervals in such a way, although we can for example retain projection factors to avoid their recomputation.

The experimental results have the \nlsat based solver performing best overall, but our new algorithm was already competitive with only a limited implementation\footnote{Referring to the lack of incrementality which meant recomputed data for different theory solver calls.}, and outperformed \nlsat on a substantial quantity of problems instances as detailed in \cref{fig:direct-comparison}.  

\subsection{Future Work}

Our next step will be to optimise the implementation of the new algorithm, in particular to have it work incrementally.  We will also consider how to optimise some of the heuristic choices required for the algorithm.  This most notably includes the variable ordering, critical for the performance of all CAD technology\footnote{We note recent work on the intersection of variable ordering between the theory and Boolean solvers \cite{Nalbach2019} and the use of machine learning to make CAD variable ordering decisions \cite{HEWBDP19, EF19, FE19a, FE20a}.}.  The new algorithm also introduces its own choices which could be made heuristically, such as whether to remove redundant intervals (Section \ref{ssec:redundancy}) and more generally which intervals to select for a covering.  

Another imminent step is to edit the algorithms and arguments so they use Lazard projection theory \cite{MPP19} in place of McCallum theory, in order to avoid the completeness issues of Section \ref{SSEC:Completeness}.  At the time of writing this was not undertaken as the additional trailing coefficients of Lazard projection made it more expensive but we note the recent work \cite{BM20} which addresses when and how these coefficients can be removed.   

Future software development will include the use of the new algorithm and NLSAT together, since the experiments showed significant data sets where one substantially outperformed the other.  It will be necessary to develop a heuristic to analyse an instance and pick the appropriate solver. 

As an application of the new algorithm, we are keen to explore whether it could provide guidance to formal proof systems.  Recent preliminary work in \cite{ADEKT20} suggests that a trace of the new algorithm provides a sequence of arguments closer to human intuition than those of the alternatives.

Other open questions include the complexity of our new algorithm, and whether more general extensions are possible.  Can the approach outlined here be adapted to more general quantifier elimination?  Can we go from ``merge intervals for existential quantifier'' to ``merge intervals for universal quantifier''?  Can we use cylindrical coverings to study all first order formulae (instead of just sets of constraints representing conjunction) and thus remove the need for the SAT solver entirely?

\subsection*{Acknowledgements}

We thank all three anonymous reviewers whose comments helped improve this paper.  We also thank Jasper Nalbach for his feedback on the paper.  The work was not funded directly by the SC$^2$ project \cite{AAB+16a} but the authors were brought together by it.

%\bibliography{references}

\end{document}